\documentclass[a4paper,11pt]{article}
\usepackage{jheppub} 

\usepackage[dvipsnames]{xcolor}
\usepackage{lipsum,amsmath,amsthm}
\usepackage{mathrsfs}
\usepackage{mathtools}
\usepackage{comment} 
\usepackage{nicematrix}
\newcommand{\be}{\begin{equation}}
\newcommand{\ee}{\end{equation}}
\newcommand{\ba}{\begin{aligned}}
\newtheorem{theorem}{Theorem}
\newtheorem*{theorem*}{Theorem}
\newtheorem{corollary}{Corollary}
\newtheorem*{corollary*}{Corollary}

\newcommand{\ea}{\end{aligned}}
\newcommand{\ri}{{\rm i}}
\newcommand{\re}{{\rm e}}
\newcommand{\cbr}[1]{\left\{ #1 \right\}}
\usepackage{tikz} 
\usepackage{tikz-cd} 
\usetikzlibrary{decorations.markings, decorations.pathreplacing, decorations.pathmorphing, shapes.misc, knots} 
\usetikzlibrary{decorations.pathreplacing} 
\usetikzlibrary{decorations.pathmorphing}
\usetikzlibrary{shapes.misc} 
\usetikzlibrary{knots} 
\newcommand{\bs}{\boldsymbol}

\newcommand{\BZ}{\mathbb{Z}}
\newcommand{\BR}{\mathbb{R}}
\newcommand{\BC}{\mathbb{C}}
\newcommand{\BP}{\mathbb{P}}
\newcommand{\pd}[2][]{\frac{\partial^{#1}}{\partial{#2}^{#1}}}
\newcommand{\ii}{\mathrm{i}}
\newcommand{\balpha}{\boldsymbol{\alpha}}
\newcommand{\br}[1]{\left( #1 \right)}
\newcommand{\sbr}[1]{\left[ #1 \right]}
\newcommand{\dd}{\mathrm{d}}
\renewcommand{\Re}{\operatorname{Re}}
\renewcommand{\Im}{\operatorname{Im}}

\DeclarePairedDelimiter{\abs}{\lvert}{\rvert}
\DeclarePairedDelimiter{\ceil}{\lceil}{\rceil}

\DeclareMathOperator{\diag}{diag}

\DeclareMathOperator{\SL}{SL}
\DeclareMathOperator{\SU}{SU}

\DeclareMathOperator{\Sym}{Sym}

\newcommand{\bm}{\boldsymbol}

\newcommand{\FNS}{F_\mathrm{NS}}

\allowdisplaybreaks

\preprint{\texttt{ZMP-HH/26-10 \\ \hspace*\fill CERN-TH-2026-110 \\ \hspace*\fill DESY-26-067}}

\title{Higher-Rank Connections and Deformed Schr\"odinger Operators
}

\author[1]{Jonah Baerman,}
\author[2,3]{Alba Grassi,}
\author[4]{Giovanni Ravazzini}
\affiliation[1]{II. Institut für Theoretische Physik, Universität Hamburg\\
Luruper Chaussee 149, 22761 Hamburg, Germany}
\affiliation[2]{Section de Math\'ematiques, Universit\'e de Gen\`eve, 1211 Gen\`eve 4, Switzerland}
\affiliation[3]{Theoretical Physics Department, CERN, 1211 Geneva 23, Switzerland}
\affiliation[4]{Deutsches Elektronen-Synchrotron DESY, Notkestr. 85, 22607 Hamburg, Germany}
\emailAdd{jonah.baerman@desy.de}
\emailAdd{alba.grassi@cern.ch}
\emailAdd{giovanni.ravazzini@desy.de}

 \abstract{We study the connection problem for a class of linear differential equations of order $N$ closely related to the Baxter equation of the quantum Toda chain. The space of solutions is $N$-dimensional and several linearly independent solutions decay at each singularity, leading to a rich structure of boundary value problems. 
 We derive the weakest quantization conditions compatible with decaying behavior at both singularities,
and formulate these conditions in terms of the associated monodromy data. In doing so, we prove the quantization conditions predicted by the topological string/spectral theory duality for a family of deformed Schr\"odinger equations. 
More generally, our results point to a hierarchy of spectral problems interpolating between the minimal conditions studied here and the maximally decaying boundary conditions of the $N$-particle quantum Toda chain.
 }

\makeatletter
\gdef\@fpheader{~}
\makeatother

\begin{document}

\maketitle
\flushbottom

\section{Introduction}

In this work we investigate the spectral properties and solutions of linear differential equations of order $N$.

When $N=2$, these reduce to standard Schr\"odinger equations, whose spectral properties have been extensively studied in the literature, see e.g.~\cite{Berezin1991} and references therein. In this case the space of solutions is two-dimensional, and one studies solutions of the equation that satisfy appropriate boundary conditions. For confining potentials, one requires square-integrable solutions, while for potentials admitting resonances one may impose the Gamow–Siegert boundary conditions. Such boundary conditions lead to a  quantization condition for the spectrum of the operator and, once this condition is imposed, the corresponding eigenfunctions are uniquely determined (up to normalization).

For higher-order equations, $N>2$, the situation is much richer because several independent solutions may decay near a given singularity, leading to a wider variety of boundary conditions and spectral problems. The class of equations we consider in this work is
\begin{equation}\label{eq:operFGP}
\left((-\ri \hbar  \partial_x)^N 
+\sum_{k=2}^{N-1} (-1)^k u_k(-\ri \hbar \partial_x)^{N-k} +\left(\Lambda^N \re^{-x}+\Lambda^N \re^x +(-1)^N u_N\right) \right)\phi(x)=0 \,,
\end{equation}
where $\Lambda,\hbar\in \BR_+$, $ x\in \BC$. For simplicity we will also consider $\{u_k\}_{k=2}^{N-1}\in \BR$. Sometimes it is  convenient to work with the variable $z=\re^x ~\in \mathbb{P}^1\setminus\{0,\infty\}$.
Equation \eqref{eq:operFGP} has two singular points, located at $x=\pm \infty$, and the space of local solutions is $N$-dimensional. Interestingly, this equation is also related to quantum integrable systems: it can be obtained from the Fourier transform of the Baxter equation for the $N$-particle quantum Toda chain \cite{Gutzwiller1980, Gutzwiller1981,PasquierGaudin1992,Sklyanin1985,Kozlowski:2010tv}.
At the same time, \eqref{eq:operFGP} also admits a natural geometric interpretation, as it arises in a special limit of the quantum mirror curves associated to toric $\rm CY_3$-folds obtained as crepant resolutions of $Y^{N,0}$ singularities \cite{Katz:1996fh,Klemm:1996bj}.

Recently, equation \eqref{eq:operFGP} and its  Fourier transform have been studied in detail in \cite{Grassi:2018bci,Francois:2025nmq} and \cite{Baerman:2025uzv}, albeit from rather different perspectives.\footnote{When $N=3$ this equation was also studied in \cite{Yan:2020kkb} in relation to spectral networks and BPS counting, see also \cite{Ito:2023zdc, Peng:2026prt} for more recent studies from the point of view of the ODE/IM correspondence. For the WKB analysis of higher-order differential equations, we refer to \cite{AokiKawaiTakei1994,AokiKawaiTakei2001,BerkNevinsRoberts1982}.
} In \cite{Grassi:2018bci,Francois:2025nmq}, the focus is on the spectral problem associated with the Fourier-transformed operator, where one constructs square-integrable solutions and derives quantization conditions using techniques from the topological string/spectral theory correspondence (TS/ST). 
In contrast, \cite{Baerman:2025uzv} interprets these equations as opers on the twice-punctured sphere $\BP^1\setminus\{0,\infty\}$ and constructs solutions to the Riemann-Hilbert and connection problems via the relation with the quantum Toda chain, which was interpreted as a realization of the analytic Langlands correspondence.

Our goal is to revisit the connection problem associated to \eqref{eq:operFGP} in light of the recent developments \cite{Baerman:2025uzv,Grassi:2018bci,Francois:2025nmq}.  As reviewed in \autoref{sec:conn matrices}, equation \eqref{eq:operFGP} admits $\lfloor N/2 \rfloor$ growing and $\lceil N/2 \rceil$ decaying solutions near each singularity.
Since several independent solutions may decay at each singularity, there are many possible ways to formulate a connection problem with $L^2(\mathbb{R})$ solutions. In this work we focus on a minimal version of the problem: we ask for the weakest condition under which at least one solution that decays near one singularity is mapped, under analytic continuation, to a solution that also decays near the other singularity. 
A different condition has also been considered in the literature in the context of integrability. In this case one typically requires maximally decaying solutions at one singularity to be mapped to maximally decaying ones at the other, leading to $N-1$ independent quantization conditions \cite{Baerman:2025uzv} making contact with \cite{Gutzwiller1980, Gutzwiller1981,PasquierGaudin1992,Sklyanin1985,Kozlowski:2010tv}. One of the outcomes of our analysis is that these two situations are part of a broader family of quantization conditions, labeled by an integer $K$, which interpolates between the minimal case studied here ($K=1$) and the integrable setting ($K=N-1$). 
We will report more on this in the future.

The main result of this work is a proof of the conjecture of \cite{Grassi:2018bci}, stated in \autoref{th:thm1} and \autoref{coroll}.\footnote{Strictly speaking the statement \eqref{eq:QCodd2} is not part of \cite{Grassi:2018bci} as in \cite{Grassi:2018bci} one only considers resonances with positive imaginary part. The condition \eqref{eq:QCodd2} corresponds to taking the negative imaginary part and it is a new result.} The proof builds on recent developments on higher-rank differential equations obtained in \cite{Baerman:2025uzv}. An alternative derivation of \cite{Grassi:2018bci} was recently proposed in \cite{Francois:2025nmq}; however, \cite{Francois:2025nmq} relies on analytic properties of the Nekrasov-Shatashvili (NS) functions in the presence of surface defects, which are currently only conjectural.

This paper is structured as follows. In \autoref{sec:mainres}, we summarize the main results. In \autoref{sec:tsst}, we discuss the relation to, and expectations from, the TS/ST correspondence. In \autoref{sec:conn matrices} we review the results of \cite{Baerman:2025uzv} and reformulate them in a way that is suitable for our purposes. In \autoref{sec:proof} we prove the theorem, and in \autoref{sec:corollproof} its corollary. We conclude  in \autoref{sec:outlook}. There are two appendices: in \autoref{app:GT}, we review the conventions for the $\mathfrak{su}(N)$ root system, and in \autoref{app:floquet} we rewrite some technical results from \cite{Baerman:2025uzv} that are relevant for making the paper self-contained.

\section*{Acknowledgments}
We would like to thank Matijn Fran\c cois, Marcos Mari\~no, Tommaso Pedroni, Paul Ryan, and J\"org Teschner for valuable discussions and collaborations on topics closely related to this work.
AG would also like to thank the DESY Theory Group for their hospitality during the initial stages of this work.
JB and GR acknowledge support by the Deutsche Forschungsgemeinschaft (DFG, German Research Foundation) under Germany's Excellence Strategy -- EXC 2121 ``Quantum Universe'' -- 390833306 and the Collaborative Research Center -- SFB 1624 ``Higher structures, moduli spaces and integrability'' -- 506632645.
The work of AG is partially supported by the Swiss National Science Foundation Grants No.~218510 and the NCCR SwissMAP.

\section{Main Results}\label{sec:mainres}
Let us consider the vectors 
\be \ba
\bs{\sigma}=\sum_{i=1}^N \sigma_i \bs e_i\,, \quad \sum_{i=1}^N \sigma_i=0\,,\\
\bs{\eta}=\sum_{i=1}^N \eta_i \bs e_i\,, \quad \sum_{i=1}^N \eta_i=0\,,
\ea\ee
which encode the monodromy data of the \(\SL(N,\mathbb{C})\) flat connection on $\mathbb{P}^1 \setminus \{0,\infty\}$ obtained from \eqref{eq:operFGP} under the change of variables $z = \re^x$. More precisely, $\sigma_j$ are defined as the eigenvalues of the
monodromy matrix around a closed curve $\gamma$ separating the two
irregular singularities on $\BP^1\setminus\{0,\infty\}$\be\label{eq:sigmamon}
M_\gamma = \operatorname{diag}\!\left(\re^{2\pi \ii\sigma_1},\ldots,\re^{2\pi \ii\sigma_N}\right) \in \SL(N,\BC)\,,
\ee
while $\eta_j$
characterize the relative normalization of the Floquet
bases of solutions defined near $z=0$ and $z=\infty$ respectively,  \cite[Prop.~4 and eq.~(5.4)]{Baerman:2025uzv},
see \autoref{app:floquet} for the details. The vectors $\bs{e}_i$ are defined in \eqref{eq:latticedef}.

Equation \eqref{eq:operFGP} is intimately related to $\SU(N)$ Seiberg--Witten theory \cite{Seiberg:1994aj, Seiberg:1994rs}, and its spectral properties nicely inherit  such an $\SU(N)$ structure. Let $\Delta_+$ denote the set of positive roots of the Lie algebra $\mathfrak{su}(N)$, and define the Weyl vector by
\be
\bs{\rho}=\frac{1}{2}\sum_{\bs{\alpha}\in \Delta_+}\bs{\alpha}\,.
\ee
We denote by $\bs{\lambda}_k$ the fundamental weights, and by $\mathcal{W}_N \cdot \bs{\lambda}_k$ their orbit under the action of the Weyl group (see \autoref{app:GT}). We then have the following results.

\begin{theorem} \label{th:thm1} \hypertarget{mainthm}
Let $N\in\mathbb{N}$, $N>1$.

\smallskip\smallskip
\noindent
\paragraph{Part I:}Assume that $N$ is even.
The differential equation \eqref{eq:operFGP} admits  non-trivial  solutions in $L^2(\mathbb{R})$  if and only if
\begin{equation}\label{eq:QCeven}
\sum_{\bs{n}\in \mathcal{W}_{N}\cdot \bs{\lambda}_{N/2}}
\frac{\exp \left(\ri\pi(2\bs{\eta}-\bs{\rho})\cdot\bs{n}\right)}
{\prod_{\bs{\alpha}\in\Delta_{+}}
\left(2\sin(\pi\,\bs{\sigma}\cdot\bs{\alpha})\right)^{(\bs{n}\cdot\bs{\alpha})^2}}
=0 \,.
\end{equation}
Such solutions necessarily exhibit super-exponential decay as $ x \to \pm\infty$.

\smallskip\smallskip
\noindent
\paragraph{Part II:}Assume that $N$ is odd.

\smallskip\smallskip
\noindent
\underline{Case 1}: The differential equation \eqref{eq:operFGP} admits  non-trivial  solutions in $L^2(\mathbb{R})$ which decay faster than any exponential at $x\to +\infty$, but may exhibit weaker decay as $x\to-\infty$, if and only if
\begin{equation}\label{eq:QCodd}
\sum_{\bs{n}\in \mathcal{W}_{N}\cdot \bs{\lambda}_{(N+1)/2}}
\frac{\exp \left(\ri\pi(2\bs{\eta}-\bs{\rho}+\bs{\sigma})\cdot\bs{n}\right)}
{\prod_{\bs{\alpha}\in\Delta_{+}}
\left(2\sin(\pi\,\bs{\sigma}\cdot\bs{\alpha})\right)^{(\bs{n}\cdot\bs{\alpha})^2}}
=0\,.
\end{equation}
\smallskip\smallskip
\noindent
\underline{Case 2}: The differential equation \eqref{eq:operFGP} admits  non-trivial  solutions in $L^2(\mathbb{R})$ which decay faster than any exponential as $x\to-\infty$, but may exhibit weaker decay as $x\to +\infty$, if and only if
\begin{equation}\label{eq:QCodd2}
\sum_{\bs{n}\in \mathcal{W}_{N}\cdot \bs{\lambda}_{(N+1)/2}}
\frac{\exp \left(\ri\pi(-2\bs{\eta}-\bs{\rho}+\bs{\sigma})\cdot\bs{n}\right)}
{\prod_{\bs{\alpha}\in\Delta_{+}}
\left(2\sin(\pi\,\bs{\sigma}\cdot\bs{\alpha})\right)^{(\bs{n}\cdot\bs{\alpha})^2}}
=0\,.
\end{equation}\\

\end{theorem}

As a corollary of \autoref{th:thm1}, we can deduce an analogous statement for the Fourier-transformed operator
\begin{equation}
\label{eq:diffm}
    \Lambda^N\left(\psi(y+\ri \hbar)+\psi(y-\ri \hbar)\right)+V_N(y) \psi(y)
    = (-1)^{N+1}u_N \psi(y)\,,
\end{equation}
with \begin{equation} \label{eq:vn}
    V_N(y) = y^N+\sum_{k=2}^{N-1} (-1)^k u_k \, y^{N-k} \, .
\end{equation}
One can also view \eqref{eq:diffm} as a deformation of the standard Schr\"odinger equation in which the usual kinetic term $p^2$ is replaced by $2\cosh p$.
For $N$ even and $u_k\in \BR$,  it was proven in \cite{Laptev:2015loa} that the difference equation \eqref{eq:diffm} has a purely discrete spectrum. By contrast, for $N$ odd a complete mathematical characterization of the spectrum is still lacking.

From \autoref{th:thm1} we can deduce the following result.
\begin{corollary} \label{coroll} 
    Let $N\in\mathbb{N}$, $N>1$.
\smallskip\smallskip
\noindent 
\paragraph{Part I:}Assume that $N$ is even. Then \eqref{eq:QCeven} holds if and only if the difference equation \eqref{eq:diffm} admits non-trivial entire solutions satisfying
\be \psi \in L^2(\mathbb{R} + \ri \alpha)\cap L^1(\mathbb{R} + \ri \alpha) \quad \text{for all} \quad \alpha \in \mathbb{R}\,,\ee
with exponential decay along every line parallel to the real axis, uniformly on every closed strip $\alpha_-\leq\alpha\leq\alpha_+$, $\alpha_\pm\in\BR$.

\smallskip\smallskip
\noindent
\paragraph{Part II:}Assume that $N$ is odd. 

 \smallskip\smallskip
\noindent
\underline{Case 1}: The difference equation \eqref{eq:diffm}
admits non-trivial entire solutions $\psi(y)$ satisfying
\be \label{eq:asycor}
\psi\in L^2(\mathbb R+\ri\alpha)
\quad \text{for all} \quad
\alpha>-\frac{\hbar}{2}\left(1-\frac{1}{N}\right),
\ee
and
\be
\psi\in L^1(\mathbb R+\ri\alpha)
\quad \text{for all} \quad
\alpha>-\frac{\hbar}{2}\left(1-\frac{3}{N}\right),
\ee
decaying and uniformly bounded on any closed strip $\alpha_-\leq\alpha\leq\alpha_+$, $\alpha_\pm>-\frac{\hbar}{2}\left( 1 - \frac{3}{N} \right)$
if and only if \eqref{eq:QCodd} holds.\footnote{The results of \cite{Francois:2025nmq}
suggest that the optimal region for $L^1$ is
$\alpha>-\frac{\hbar}{2}\left(1-\frac{2}{N}\right)$. It is possible to prove this sharper bound from \cite{DiNezzaPalatucciValdinoci2012}, but we will not need it here.}

 \smallskip\smallskip
\noindent
\underline{Case 2}:
The difference equation \eqref{eq:diffm}
admits non-trivial entire solutions $\psi(y)$ satisfying
\be \label{eq:asycor2}
\psi\in L^2(\mathbb R+\ri\alpha)
\quad \text{for all} \quad
\alpha<\frac{\hbar}{2}\left(1-\frac{1}{N}\right),
\ee
and
\be
\psi\in L^1(\mathbb R+\ri\alpha)
\quad \text{for all} \quad
\alpha<\frac{\hbar}{2}\left(1-\frac{3}{N}\right),
\ee
decaying and uniformly bounded on any closed strip $\alpha_-\leq\alpha\leq\alpha_+$, $\alpha_\pm<\frac{\hbar}{2}\left( 1 - \frac{3}{N} \right)$
if and only if \eqref{eq:QCodd2} holds.\footnote{The results of \cite{Francois:2025nmq}
suggest that the optimal region for $L^1$ is
$\alpha<\frac{\hbar}{2}\left(1-\frac{2}{N}\right)$. It is be possible to prove this sharper bound from \cite{DiNezzaPalatucciValdinoci2012}, but we will not need it here.}

\end{corollary}

The conditions on uniform boundedness/decay are purely technical, as they are sufficient to allow for shifts of integration contours of the Fourier transform along the imaginary direction.
Given the results of~\cite{Francois:2025nmq} as well as~\cite{Gutzwiller1980,Gutzwiller1981,PasquierGaudin1992}, we expect that these conditions may be dropped, and follow automatically from equation~\eqref{eq:diffm}.

\section{Relation to TS/ST Correspondence}\label{sec:tsst}

The topological string/spectral theory (TS/ST) correspondence is an exact duality relating topological string theory on local Calabi-Yau threefolds to the spectral theory of quantum mirror curves. This correspondence naturally decomposes into two complementary sectors: a closed string sector and an open string sector, originally formulated in \cite{Grassi:2014zfa,Codesido:2015dia} and \cite{Marino:2016rsq,Marino:2017gyg,Francois:2025wwd,Francois:2025nmq}, respectively. 
In the closed sector, the fundamental objects on the string  side are the closed topological string partition functions, which are conjecturally equivalent to spectral quantities encoding the spectrum of the associated quantum operator, such as Fredholm determinants or spectral traces. In contrast, the open sector provides a direct bridge between open topological string partition functions and the exact eigenfunctions of the corresponding quantum spectral problem.
This duality has been extensively tested in a wide variety of examples, both numerically and analytically, and some of its aspects have been demonstrated rigorously for instance in \cite{Babu:2024hxg,Kashaev:2015kha, Kashaev:2015wia,Bonelli:2016idi,Laptev:2015loax,Gavrylenko:2023ewx}. 
From this perspective, the results presented in this work provide a mathematical derivation of certain predictions of the TS/ST correspondence. 

Indeed, the quantization conditions \eqref{eq:QCeven} and \eqref{eq:QCodd} were originally obtained in \cite{Grassi:2018bci} as predictions arising from the closed sector of the TS/ST correspondence. To make contact with \cite{Grassi:2018bci} we also have to relate $\eta_i$ to the NS free energy, i.e.~the Yang–Yang function of the closed quantum Toda chain.
This was done in
\cite{ns,Kozlowski:2010tv,Nekrasov:2011bc, Baerman:2025uzv} where it was found that
\begin{equation}\label{eq:eta-NS}
    2\pi\ii\bm\eta = \frac{\ii}{\hbar}\pd{\bm a}\FNS  + \pi\ii\bm\rho\,,\qquad {\bs a}=-\ri \hbar {\bs \sigma}\,,
\end{equation}
where $ \FNS$ is the Yang–Yang function, or  NS free energy.
At small $\Lambda$ it behaves as
\be \ba 
 \frac{\ri}{\hbar} \partial_{\bs{a}} F_{\mathrm{NS}}
    =
    - \sum_{\balpha \in \Delta_+} 
    \sbr{\ri \, \frac{\pi}{2}
    + \ri \, 2 \br{\balpha \cdot \frac{\bs{a}}{\hbar}} \log\!{\br{\frac{\Lambda}{\hbar}}}
    + \log\!{\br{\frac{
    \Gamma\!{\br{1 - \ri \br{\balpha \cdot \frac{\bs{a}}{\hbar}}}}}
    {\Gamma\!{\br{1 + \ri \br{\balpha \cdot \frac{\bs{a}}{\hbar}}}}}
    }}} \balpha  +\mathcal{O}(\Lambda^N)\,.
\ea\ee

The relation  \eqref{eq:eta-NS} was proven in \cite{Kozlowski:2010tv, Baerman:2025uzv} using the representation of $\FNS$ in terms of a nonlinear integral equation (NLIE). On the other hand in \cite{Grassi:2018bci}, a different representation in terms of Young diagrams was used. These two expressions are only conjecturally equivalent, see e.g.~\cite{ns,Meneghelli:2013tia,Kozlowski:2010tv} for more details and a discussion on the overlapping region of these representations.
In addition, $\sigma_i$ and $\eta_i$ are  defined at generic points in the moduli space as we explain in \autoref{app:floquet}, whereas the identification \eqref{eq:eta-NS} is expected to hold  in a smaller region of the moduli space, where the NLIE and/or the Young-diagram representation converges.

More recently, in \cite{Francois:2025nmq}, the conjecture of \cite{Grassi:2018bci} was made more concrete through the construction of explicit eigenfunctions of \eqref{eq:diffm}. The construction is  rigorous; however it relies  on certain analytic properties of the NS functions in the presence of surface defects, that are presently known only conjecturally. Let us briefly review some of the results in \cite{Francois:2025nmq} which are a useful guideline for our purposes.
First, in \cite[eqs.~(2.25), (2.31)]{Francois:2025nmq} one constructs solutions to \eqref{eq:diffm} which are entire in the complex $y$-plane for generic values of $u_k$. 
Such functions, however, are not square-integrable: they decay in one direction, while in the other they oscillate with growing amplitude.
Nevertheless, by imposing the quantization conditions (\ref{eq:QCeven}, \ref{eq:QCodd}, \ref{eq:QCodd2}), one obtains physical eigenfunctions with the following decay properties:  
\begin{itemize}
\item If $N$ is even, we have bound states and \eqref{eq:QCeven} gives eigenfunctions that are in $L^2$ on any line parallel to the real axis, and along such lines they decay exponentially at $\pm\infty$ as
\be\label{eq:dec1}
\psi(y)\sim\exp\left(-\frac{\pi}{\hbar}\abs{{\rm Re}(y)}\right), \quad {\rm Im}(y)\quad \text{fixed}.
\ee
\item
If $N$ is odd, we have resonance states and it is well known that resonances always occur in complex-conjugate pairs  \cite{complexdila,Caliceti, Caliceti1983, Maioli}. 

Independently of which resonance is considered, as $y\to +\infty$, along any line parallel to the real axis the eigenfunctions decay exponentially as
\be\label{eq:dec2}
\psi(y)\sim\exp\left(-\frac{\pi}{\hbar}\,{\rm Re}(y)\right), \quad {\rm Im}(y)\quad \text{fixed}, \quad {\rm Re}(y)\to \infty\,.
\ee
As $y \to -\infty$, the decay is instead only polynomial and depends on which member of the complex-conjugate pair is selected.

\begin{itemize}
    \item 
    Let us first consider the case in which the resonance has positive imaginary part, as in \cite{Grassi:2018bci,Francois:2025nmq}. After imposing \eqref{eq:QCodd}, the eigenfunction becomes square-integrable along any line parallel to the real axis provided that
\be
{\rm Im}(y)> -\frac{\hbar}{2}\left({1}-{1\over N}\right).
\ee
Along such lines, the asymptotic behavior is
\be\label{eq:dec3}
\psi(y)\sim|{\rm Re}(y)|^{-\frac{N}{2}-\frac{N}{\hbar}\,{\rm Im}(y)}, \quad {\rm Im}(y)\quad \text{fixed}, \quad  {\rm Re}(y)\to -\infty\,.
\ee
\item Consider instead the complex-conjugate resonance with negative imaginary part. This case was not considered explicitly in \cite{Grassi:2018bci,Francois:2025nmq}. However, it corresponds to taking the complex conjugate off-shell solution (i.e.~$\overline{\psi\left(\overline{y},\overline{u}\right)}$) and, as we discuss below, to imposing the condition \eqref{eq:QCodd2}.   In this case, the eigenfunction is square-integrable along any line parallel to the real axis provided that
\be
{\rm Im}(y) < \frac{\hbar}{2}\left(1-{1\over N}\right),
\ee
and along such lines the decay is
\be\label{eq:dec4}
\psi(y)\sim|{\rm Re}(y)|^{-\frac{N}{2}+\frac{N}{\hbar}\,{\rm Im}(y)}, \quad {\rm Im}(y)\quad \text{fixed},\quad  {\rm Re}(y)\to -\infty\,.
\ee
\end{itemize}
\end{itemize}
The relation between \eqref{eq:diffm} and \eqref{eq:operFGP} is given by a simple Fourier transform 
\be \label{eq:Fourier-conventions}
\ba \ri \hbar \partial_y \quad &\to \quad x\,,\\
 y \quad &\to \quad -\ri \hbar \partial_x\,.\\
\ea
\ee
 At the level of solutions this reads
\be \phi(x)={1\over \sqrt{2\pi\hbar}}\int_{\BR}\dd y\, \re^{{\ri\over \hbar} x y } \psi(y)\,.\ee
It is then useful to understand how  (\ref{eq:dec1}, \ref{eq:dec2}, \ref{eq:dec3}, \ref{eq:dec4}) transform after applying the Fourier transform.
\begin{itemize}
    \item Let us first consider the case in which $N$ is even. Since $\psi(y)$ is entire and decays as \eqref{eq:dec1}, by the Riemann-Lebesgue lemma it follows that the Fourier transform exists, is bounded and vanishes at infinity. 
In addition, we have 
\begin{equation}
\phi(x)=\frac{1}{\sqrt{2\pi\hbar}}\int_{\mathbb{R}+ \ii\epsilon} \dd y\, 
\re^{\frac{\ii}{\hbar}xy}\psi(y)
=\frac{\re^{-\frac{\epsilon x}{\hbar}}}{\sqrt{2\pi\hbar}}\int_{\mathbb{R}} \dd y\, 
\re^{\frac{\ri}{\hbar}xy}\psi(y+\ri\epsilon)\,,
\qquad \forall \epsilon\in\mathbb{R}\,.
\end{equation}
In particular, because of \eqref{eq:dec1}, this means that  $\phi(x)$ decays at $\pm\infty$ faster than
\begin{equation}
\re^{-\frac{\epsilon |\mathrm{Re}(x)|}{\hbar}}, 
\qquad \forall \epsilon>0 \,.
\end{equation}
As we will see later this is in agreement with the double exponential decay of the solutions  of \eqref{eq:operFGP} near its singularities \cite{Baerman:2025uzv}.

\item Let us now consider the case in which $N$ is odd. As before, entireness and conditions (\ref{eq:dec2}, \ref{eq:dec3},  \ref{eq:dec4}) ensure that the Fourier transform exists, is bounded and vanishes at infinity.\footnote{The case $N=3$ is very slightly more involved, but the outcome is the same.}
\begin{itemize}
    \item Let us first examine the case of resonances with positive imaginary part.
Because of \eqref{eq:dec3}, the integration contour in the Fourier transform can be shifted by $\ri\epsilon$ only as long as $\epsilon >-\frac{\hbar}{2}\left({1}-{3\over N}\right)$. Hence, as $x\to +\infty$, we still find that $\phi(x)$ decays faster than
\begin{equation}
\re^{-\frac{\epsilon |\mathrm{Re}(x)|}{\hbar}}, 
\qquad \forall \epsilon>0 \,.
\end{equation}
On the other hand, the same argument cannot be applied as $x\to -\infty$ because of the lower bound on $\epsilon$.
We therefore expect a weaker decay along this direction.

\item For resonances with negative imaginary part, the result is similar. In particular, because of \eqref{eq:dec4}, the integration contour in the Fourier transform can be shifted by $\ri\epsilon$ only as long as $\epsilon <\frac{\hbar}{2}\left(1-{3\over N}\right)$. Hence, as $x\to -\infty$, we still find that $\phi(x)$ decays faster than
\begin{equation}
e^{-\frac{\epsilon |\mathrm{Re}(x)|}{\hbar}}, 
\qquad \forall \epsilon>0 \,.
\end{equation}
On the other hand, the same argument cannot be applied as $x\to +\infty$ because of the upper bound on $\epsilon$,  we therefore expect a weaker decay along this direction.
\end{itemize}
\end{itemize}

Let us note that \eqref{eq:diffm} is closely related to the Baxter equation of the Toda lattice studied in \cite{Baerman:2025uzv}. The latter reads 
\begin{equation} \label{eq:baxter}
(\mathrm{i}\Lambda)^N Q(y+\mathrm{i} \hbar)+(-\mathrm{i}\Lambda)^N Q(y-\mathrm{i} \hbar)-V_N(y)Q(y)=(-1)^Nu_N Q(y)\,.
\end{equation}
Hence, when $N \equiv 2 \ (\mathrm{mod}\ 4)$ the two equations agree, while in the other cases they can be related by a non-trivial transformation of $Q$, see e.g. \cite[eq.~(3.3)]{Francois:2025nmq}.

For instance, if $N \equiv 0 \ (\mathrm{mod}\ 4)$, the Baxter equation \eqref{eq:baxter} can be brought to the form \eqref{eq:diffm}, at the cost of inverting the potential.
On the differential equation side, this discrepancy can be absorbed by a rotation of the variable $z$. Indeed, the Fourier transform of \eqref{eq:baxter}, taken according to~\eqref{eq:Fourier-conventions}, takes the form
\begin{equation}\label{eq:operBRT}
\left((-\mathrm{i} \hbar z \partial_z)^N
+\sum_{k=2}^{N-1} (-1)^k u_k(-\mathrm{i} \hbar z \partial_z)^{N-k}
- (\mathrm{i} \Lambda)^N z-(-\mathrm{i} \Lambda)^N \frac{1}{z}+(-1)^N u_N \right)\chi(z)=0 \,,
\end{equation}
where again $z=\re^{x}$. Therefore, in order to recover \eqref{eq:operFGP}, one needs to perform the rotation
\begin{equation} \label{eq:z-rotation}
z ~\to~ (-\ri)^{N+2} z\,.
\end{equation}
In this work, we adopt the conventions of \eqref{eq:operFGP} and \eqref{eq:diffm}, as they are more convenient for our purposes. Alternative conventions could also be used, but they would lead to different asymptotic behaviors, see also \cite[eq.~(3.3)]{Francois:2025nmq}.

Let us conclude this section by noting that the left hand side of \eqref{eq:QCeven} and \eqref{eq:QCodd} are, up to a $u_N$-independent normalization,  the Fredholm determinants of the associated spectral problem; see \cite[App.~A]{Grassi:2019coc}.

\section{Bases of Solutions and Connection Matrix}\label{sec:conn matrices}
In this section we avail ourselves of the results of \cite{Baerman:2025uzv} in order to define two bases of solutions $\boldsymbol{\phi}^{(0,\infty)}(z)$ to equation (\ref{eq:operFGP}) having a prescribed asymptotic behavior at $z=0,\infty$ respectively. Furthermore, we review how to relate them by a connection matrix $E$, which ultimately determines the quantization conditions (\ref{eq:QCeven}, \ref{eq:QCodd}, \ref{eq:QCodd2}).

The  equation~\eqref{eq:operBRT} admits two bases of solutions $\chi^{(\infty)}_k$ and $\chi^{(0)}_k$ ($k=1,\dots,N)$ which admit the following asymptotic series expansions as $z\to0,\infty$ respectively
\begin{subequations} \label{eq:chi-asymptotics}
\begin{align}
    \chi_k^{(0)}(z) &\sim \re^{-\alpha_{-k}N\frac{\Lambda}{\hbar}z^{-1/N}}z^\frac{N-1}{2N}\left( 1 + {\cal O}\bigl( z^{1/N} \bigr) \right), \\
    \chi_k^{(\infty)}(z) &\sim \re^{-\alpha_kN\frac{\Lambda}{\hbar}z^{1/N}}z^{-\frac{N-1}{2N}}\left( 1 + {\cal O}\bigl(z^{-1/N}\bigr) \right),
\end{align}
\end{subequations}
where $\alpha_k=\re^{\frac{2\pi\ii}{N}k}$ is a root of unity. If ${\rm Re}(\alpha_{\pm k})>0$, we say that the decay/growth in \eqref{eq:chi-asymptotics} is super-exponential since  $z=\re^x$.

Let us consider a distinguished pair of solutions $\chi^{(0,\infty)}_\mathrm{max}\equiv\chi^{(0,\infty)}_0$, which have maximal decay, corresponding to $\alpha_0=1$ at their respective punctures.
In this case, the asymptotics~\eqref{eq:chi-asymptotics} are valid in a sector centered at $z=\infty$ with opening angle $\abs{\arg z}<\pi(N+1)$ (see \cite[Prop.~5]{Baerman:2025uzv} for the precise definition).
A full basis of solutions can be defined by analytic continuation of $\chi^{(0,\infty)}_\mathrm{max}$ as\footnote{Here we reuse the notation $\chi^{(0,\infty)}_k$ to refer to a slightly different basis than the resummation of~\eqref{eq:chi-asymptotics}. The reader should keep in mind that \eqref{eq:chi-asymptotics} $\sim$ \eqref{eq:chi-basis-def} only for $\theta_0=0$, and only for the solutions which have $\abs{\theta_0+2\pi k}<\pi(N+1)$. Taking $\theta_0\neq0$ amounts to a shift of the label, while solutions with $\abs{\theta_0+2\pi k}>\pi(N+1)$ generically have maximal growth due to Stokes phenomena.}
\begin{equation} \label{eq:chi-basis-def}
    \chi^{(0,\infty)}_k(z) \equiv \chi^{(0,\infty)}_\mathrm{max}\bigl( z\re^{\ii(\theta_0+2\pi k)} \bigr)\,,
\end{equation}
with an arbitrary starting angle $\theta_0\in2\pi\BZ$.
For a suitable choice of $\theta_0$ it is straightforward to verify that $\abs{\theta_0+2\pi j}<\pi(N+1)$ for all $j=1,\dots,N$, such that the asymptotics~\eqref{eq:chi-asymptotics} remain valid, and linear independence is ensured.
Since applying a monodromy matrix (which is invertible) to this basis of solutions simply shifts $\theta_0\to\theta_0+2\pi$, linear independence is guaranteed for any choice of $\theta_0\in2\pi\BZ$.

In order to obtain solutions to  equation~\eqref{eq:operFGP}, we need to perform the additional rotation~\eqref{eq:z-rotation} of the $z$ coordinate.
That is, given a solution $\chi(z)$ of equation~\eqref{eq:operBRT}
\begin{equation}
    \phi(z) = \chi\bigl( z\re^{-\ii\frac{\pi}{2}(N+2)} \bigr)\,,
\end{equation}
is a solution of equation~\eqref{eq:operFGP}.
To construct a basis for the modified equation~\eqref{eq:operFGP} analogously to~\eqref{eq:chi-basis-def}, we can therefore simply add this shift to the starting angle $\theta_0\to\varphi_0$, such that
\begin{equation}
    \varphi_0 + \frac{\pi}{2}(N+2) \in 2\pi\BZ\,.
\end{equation}
Specifically we define
\begin{equation} \label{eq:phi-basis-def}
    \phi^{(0,\infty)}_j(z) = \chi^{(0,\infty)}_\mathrm{max}\bigl( z\re^{\ii(\varphi_0+2\pi j)} \bigr)\,,\qquad \varphi_0 = -\frac{\pi}{2}(N+2)\,.
\end{equation}
This choice of $\varphi_0$ is such that $\phi^{(0,\infty)}_j$ is decaying at $0,\infty$ respectively for $j=1,\dots,\ceil{\tfrac{N}{2}}$. 
The asymptotics of these solutions are depicted in~\autoref{fig:roots-of-unity-ordered} for some low-lying examples.
For $N$ odd, we remark that there is a distinguished basis element $\phi^{(0,\infty)}_{\frac{N+1}{2}}$ which lies on the boundary between the growing and decaying regions, and has asymptotics
\begin{subequations} \label{eq:oscillating-sol}
\begin{align}
    \phi^{(0)}_\frac{N+1}{2}(z) &\sim \re^{\ii N\frac{\Lambda}{\hbar}z^{-1/N}}z^\frac{N-1}{2N}\,,\qquad\;\;\, z\to0\,, \\
    \phi^{(\infty)}_\frac{N+1}{2}(z) &\sim \re^{-\ii N\frac{\Lambda}{\hbar}z^{1/N}}z^{-\frac{N-1}{2N}}\,,\qquad z\to\infty\,.
\end{align}
\end{subequations}
Crucially, it still decays at $0,\infty$ respectively, but now only polynomially instead of exponentially.

\begin{figure}[ht!]
\centering

\begin{tikzpicture}

\fill [red!5] (-3,-3) rectangle (0,3);
\fill [red!5] (5,-3) rectangle (8,3);
\fill [red!5] (-3,-11) rectangle (0,-5);
\fill [red!5] (5,-11) rectangle (8,-5);
\fill [green!5] (0,-3) rectangle (3,3);
\fill [green!5] (8,-3) rectangle (11,3);
\fill [green!5] (0,-11) rectangle (3,-5);
\fill [green!5] (8,-11) rectangle (11,-5);

\begin{scope}
    \clip (0,0) -- (-3,-3) -- (0,-3) -- (0,0);
    \foreach \i in {1,...,20}
        \draw [very thin] (0,-6+0.3*\i) -- (-3,-3+0.3*\i);
\end{scope}

\begin{scope}
    \clip (8,0) -- (5,-3*0.726543) -- (5,-3) -- (8,-3) -- (8,0);
    \foreach \i in {1,...,20}
        \draw [very thin] (8,-6+0.3*\i) -- (5,-3+0.3*\i);
\end{scope}

\begin{scope}
    \clip (0,-8) -- (-3,-8-3*0.57735) -- (-3,-11) -- (0,-11) -- (0,-8);
    \foreach \i in {1,...,20}
        \draw [very thin] (0,-14+0.3*\i) -- (-3,-11+0.3*\i);
\end{scope}

\begin{scope}
    \clip (8,-8) -- (5,-8-3*0.481575) -- (5,-11) -- (8,-11) -- (8,-8);
    \foreach \i in {1,...,20}
        \draw [very thin] (8,-14+0.3*\i) -- (5,-11+0.3*\i);
\end{scope}

\draw (0,0) -- (-3,-3);
\draw (8,0) -- (5,-3*0.726543);
\draw (0,-8) -- (-3,-8-3*0.57735);
\draw (8,-8) -- (5,-8-3*0.481575);

\draw [->] (-3,0) -- (3,0) node [right] {$\Re \alpha_k$};
\draw [->] (5,0) -- (11,0) node [right] {$\Re \alpha_k$};
\draw [->] (-3,-8) -- (3,-8) node [right] {$\Re \alpha_k$};
\draw [->] (5,-8) -- (11,-8) node [right] {$\Re \alpha_k$};
\draw [->] (0,-3) -- (0,3) node [above] {$\Im \alpha_k$};
\draw [->] (8,-3) -- (8,3) node [above] {$\Im \alpha_k$};
\draw [->] (0,-11) -- (0,-5) node [above] {$\Im \alpha_k$};
\draw [->] (8,-11) -- (8,-5) node [above] {$\Im \alpha_k$};

\draw [thick] circle (2cm);
\draw [thick] (8,0) circle (2cm);
\draw [thick] (0,-8) circle (2cm);
\draw [thick] (8,-8) circle (2cm);

\foreach \i in {1,...,4}
    \filldraw [thick, fill=white] (0,0) ++(-135+90*\i:2) circle (7pt) node {\i};
    
\foreach \i in {1,...,5}
    \filldraw [thick, fill=white] (8,0) ++(90+72*\i-3*72:2) circle (7pt) node {\i};

\foreach \i in {1,...,6}
    \filldraw [thick, fill=white] (0,-8) ++(60*\i-120:2) circle (7pt) node {\i};

\foreach \i in {1,...,7}
    \filldraw [thick, fill=white] (8,-8) ++(90+360/7*\i-360/7*4:2) circle (7pt) node {\i};

\draw [thick, ->] +(-45:1.5) arc (-45:5*45:1.5);
\draw [thick, ->] (8,0)+(90-2*72:1.5) arc (90-2*72:90+2*72:1.5);
\draw [thick, ->] (0,-8)+(-60:1.5) arc (-60:4*60:1.5);
\draw [thick, ->] (8,-8)+(90-360/7*3:1.5) arc (90-360/7*3:90+360/7*3:1.5);

\node at (0,-3.5) {$N=4$};
\node at (8,-3.5) {$N=5$};
\node at (0,-11.5) {$N=6$};
\node at (8,-11.5) {$N=7$};

\end{tikzpicture}
\caption{The families of solutions $\phi^{(0,\infty)}_k$ of equation~\eqref{eq:operFGP} defined through equation~\eqref{eq:phi-basis-def} have asymptotics given by $\phi^{(0)}_k(z)\sim\mathrm{const}\times \re^{-\alpha_k^{-1}N\tfrac{\Lambda}{\hbar}z^{-1/N}}z^\frac{N-1}{2N}$ and \mbox{$\phi^{(\infty)}_k(z)\sim \mathrm{const}\times \re^{-\alpha_kN\tfrac{\Lambda}{\hbar}z^{1/N}}z^{-\frac{N-1}{2N}}$} as $z\to0,\infty$ respectively, where $\alpha_k = \re^{\ii(\varphi_0+2\pi k)/N}$ is depicted above. $\Re\alpha_k<0$ corresponds to exponentially growing solutions, and $\Re\alpha_k>0$ to exponentially decaying ones. The special case $\alpha_k=\ii$ appearing for odd $N$ only decays polynomially, while the dashed region has $\abs{\varphi_0+2\pi k}>\pi(N+1)$, such that the asymptotics~\eqref{eq:chi-asymptotics} are no longer valid.} 
\label{fig:roots-of-unity-ordered}
\end{figure}

 In \cite{Baerman:2025uzv} two bases of Floquet solutions to equation (\ref{eq:operBRT}), denoted $F^{(0,\infty)}_i(z)$, were introduced. These are bases of solutions whose monodromy around the cylinder $\BC \setminus\{0,\infty\}$ is represented by the diagonal matrix (\ref{eq:sigmamon}); we recall their construction in~\autoref{app:floquet}. The maximally decaying solutions to (\ref{eq:operBRT}) at $z=0,\infty$ can be expanded in the bases of Floquet solutions as
\begin{equation} \label{eq:chi-max-Floquet}
    \chi_\mathrm{max}^{(0,\infty)}(z) = \sum_{i=1}^N c_iF^{(0,\infty)}_i(z)\,.
\end{equation}
respectively. The precise coefficients $c_i=c_i(\bm\sigma)$ were computed in~\cite{Baerman:2025uzv} using an explicit construction of $\chi_\mathrm{max}^{(0,\infty)}$, but they are not essential here. 
The only crucial point is that they are the same regardless of the choice of labelling $0,\infty$ in (\ref{eq:chi-max-Floquet}). 

In analogy to~\eqref{eq:phi-basis-def}, we can define a pair of Floquet bases for  equation~\eqref{eq:operFGP} as
\begin{equation}
    G^{(0,\infty)}_i(z) = c_iF^{(0,\infty)}_i\bigl( z\re^{\ii(\varphi_0+2\pi)} \bigr)\,.
\end{equation}
Since their ratio does not depend on $z$, we still have
\begin{equation}
    \frac{G^{(0)}_i(z)}{G^{(\infty)}_i(z)} = \frac{F^{(0)}_i(z)}{F^{(\infty)}_i(z)} = \frac{Q^+_{\bm \tau}(-\ri\hbar \sigma_i)}{Q^-_{\bm\tau}(-\ri\hbar \sigma_i)} \eqqcolon\zeta_i\,,
\end{equation}
where we let $\zeta_i\coloneq\re^{2\pi\ii\eta_i}$ and $Q^\pm_{\bm\tau}$ is defined in \autoref{app:floquet}. 
Using the definition~\eqref{eq:phi-basis-def}, the expansion~\eqref{eq:chi-max-Floquet}, and the fact that the $G^{(0,\infty)}_i$ have diagonal monodromy,  we conclude from \cite[Lemma 6]{Baerman:2025uzv} that the basis elements $\phi_j^{(0,\infty)}(z)$ introduced in (\ref{eq:phi-basis-def}) can be expanded as
\begin{equation}
    \phi^{(0,\infty)}_j(z) = \sum_{i=1}^N G^{(0,\infty)}_i (z)\Sigma_i^{j-1}\,,
\end{equation}
where $\Sigma_i = \re^{2\pi\ii\sigma_i}$ and $\sigma_i$ is the Floquet exponent of $F^{(0,\infty)}_i$, see \autoref{app:floquet}.
The connection matrix $E$ satisfying $\bm\phi^{(0)} = \bm\phi^{(\infty)}E$, where $\bm\phi^{(0,\infty)}\coloneqq (\phi_1^{(0,\infty)}(z),\dots,\phi_N^{(0,\infty)}(z))$ can therefore be obtained as \cite{Baerman:2025uzv}
\begin{equation}\label{eq:Edef}
  \boxed{ E = V^{-1}TV }  
\end{equation}
where $T=\diag(\zeta_1,\dots,\zeta_N)$ and $V=V(\Sigma_1,\dots,\Sigma_N)$ is a Vandermonde matrix in the variables $\Sigma_1,\dots,\Sigma_N$.
We note that the formula~\eqref{eq:Edef} does not depend in any way on the choice of $\varphi_0$, which only affects the interpretation of each basis element.
The choice in~\eqref{eq:phi-basis-def} is therefore largely arbitrary, and a different choice of $\varphi_0$ would simply shift the position of the decaying elements in the index set $\{1,\dots,N\}$.
Since the quantization conditions are basis independent, this gives the same result.

\section{Proof of Main Theorem}\label{sec:proof}

Having introduced all the relevant definitions, we can now proceed with the proof of \autoref{th:thm1}, which consists of three steps
\begin{itemize}
    \item Compute the determinant of the bottom-left, $M\times M$ submatrix $E[\swarrow]$ of the connection matrix $E$, with $M= \ceil{N/2}$. 
    \item  Show that the decay properties of the solutions to (\ref{eq:operFGP}) required by \autoref{th:thm1} are equivalent  to the condition $\det E[\swarrow]=0$.
    \item Argue that $\det E[\swarrow]$ is given by (\ref{eq:QCeven}, \ref{eq:QCodd}, \ref{eq:QCodd2}) depending on the parity of $N$ and, for $N$ odd, on whether solutions are required to decay exponentially at $\pm\infty$, respectively.\footnote{To obtain (\ref{eq:QCodd2}) we will in fact consider the bottom-left, $M\times M$ submatrix of $E^{-1}$, whose determinant can be computed by simply replacing $\eta_k \rightarrow -\eta_k$ in the final formula (\ref{det-formula-1}).} 
\end{itemize}
\subsection{The Bottom-Left Minor of the Connection Matrix}
In this subsection, we compute the determinant of the $M \times M$ submatrix located in the bottom-left corner of the matrix $E$ (\ref{eq:Edef}), where $M=\ceil{N/2}$, which we denote by $E[\swarrow]$. 

To compute $\det E[{\swarrow}]$, let us first fix some useful notation. 
Let $L=\{1,\dots,N\}$, $S \subset L$ with $|S|=M$, $M _1=\{1,\dots,M\}$, $M_2=\{N-M+1,\dots,N\}$. Let $S^c = L\setminus S$ be the complement of $S$ in $L$ and $\sum S = \sum_{i \in S}i$ be the sum of all elements of the subset $S$. For brevity, let $A_{IJ}$ be the matrix obtained by selecting the rows labeled by $I$ and the columns labeled by $J$ of a bigger matrix $A$.\footnote{To be precise, let $I=\{I_1,\dots,I_M\}$ and  $J=\{J_1,\dots,J_M\}$. Then one has $(A_{IJ})_{ij} = A_{I_i J_j}$.}
Finally we denote by $\Sym(S \times S^c)$ the set $\{(i,j) \,|\, i \in S, j\in S^c \lor i \in S^c, j\in S\}$.

Recall the expression (\ref{eq:Edef}) for the connection matrix $E$. By the Cauchy-Binet theorem (see e.g.~\cite[Subsection 0.8.7]{Horn_Johnson_1985}) %
we have that
\begin{equation}\label{eq:Cauchy-Binet}
    \det E[\swarrow]= \sum_{S\subset L} \det ((V^{-1})_{M_2 S}) \det ((TV)_{SM_1})\,.
\end{equation}
Observing that $M_1$ selects the first $M$ columns of the Vandermonde matrix, we can use the Vandermonde determinant formula to find
\begin{equation}
    \det ((TV)_{SM_1}) = \prod_{i\in S} \zeta_i\prod_{\substack{i,j \in S \\ i>j}}(\Sigma_i-\Sigma_j)\,.
\end{equation}
By the Jacobi complementary minor formula (see e.g.~\cite[Lemma A.1 (e)]{CaraccioloSokalSportiello2013}), we have that
\begin{equation}
    \det( (V^{-1})_{M_2 S}) = \frac{(-1)^{ \sum S +  \sum M_2}}{\det V} \det (V_{S^cM_2^c} )\,,
\end{equation} 
where, again using the Vandermonde determinant formula, we can write the latter factor as
\begin{equation}
    \det( V_{S^cM_2^c}) = \prod_{\substack{i,j \in S^c \\ i>j}}(\Sigma_i-\Sigma_j)\,.
\end{equation}
Putting everything together, we find
\begin{align}\label{det-formula-1}
    \det E[\swarrow] &= \sum_{S\subset L}(-1)^{ \sum S + \sum M_2}\prod_{i\in S} \zeta_i\frac{ \prod_{i,j \in S^c;i>j}(\Sigma_i-\Sigma_j) \prod_{i,j \in S; i>j}(\Sigma_i-\Sigma_j)}{\prod_{i,j \in L;i>j}(\Sigma_i-\Sigma_j)} \\ \nonumber 
  \label{detE21-1}  &=  \boxed{\sum_{S\subset L}(-1)^{ \sum S +  \sum M_2}\prod_{i\in S} \zeta_i\prod_{\substack{(i,j) \in \Sym(S\times S^c) \\ i>j}}\frac{1}{\Sigma_i-\Sigma_j}}
\end{align}
where in the last line we simply observe that the terms involving differences of $\Sigma_i-\Sigma_j$ with indices $i,j$ both in $S$ or $S^c$ simplify.

For completeness, let us observe that we can reach a more synthetic formula for (\ref{det-formula-1}). We can rewrite
\begin{equation}
\prod_{\substack{(i,j) \in \Sym(S\times S^c) \\ i>j}}\frac{1}{\Sigma_i-\Sigma_j}=  (-1)^{r}\prod_{(i,j)\in S\times S^c}\frac{1}{\Sigma_i-\Sigma_j}\,,
\end{equation}
where $r$ is given by
\begin{equation}
    r = \abs{\{(i,j) \in S \times S^c \,|\, i<j\}}\,.
\end{equation}
If we fix $i \in S$, then the number $r_i $ of $j \in S^c$ such that $j > i$, is the total number of indices $i' \in L$  such that $i'>i$, minus the number of indices  $i'' \in S$ such that $i'' >i$, namely 
\begin{equation}
r_i = N-i - \abs{\{k \in S \,|\, k>i\}}\,.
\end{equation}
Then, observing that the cardinality of $\{k \in S \,|\, k>i\}$ takes the values $0,\dots,M-1$ as $i$ varies in $S$, we find
\begin{equation}
    r = \sum_{i\in S}r_i = - \sum  S 
    +\frac{1}{2} M (2 N-M+1)\,.
\end{equation}
Observing finally that 
\begin{equation}
    \sum M_2 = \sum_{k=N-M+1}^{N}k=\frac{M(2N-M+1)}{2}\,,
\end{equation}
we find that $\sum S +\sum M_2 +r = M (2 N - M + 1)$, which is always even. Hence we can express (\ref{det-formula-1}) by
\begin{equation}
  \det E[\swarrow] = \sum_{S\subset L}\prod_{i\in S} \zeta_i\prod_{i\in S, j \in S^c}\frac{1}{\Sigma_i-\Sigma_j}\,.
\end{equation}

\subsection{Square-Integrable Solutions for \texorpdfstring{$N$}{N} Even}\label{sec:deteven}

When $N$ is even, around each singularity, the basis of solutions to (\ref{eq:operFGP}) 
\be \boldsymbol{\phi}^{(0,\infty)}\coloneqq (\phi_1(z),\dots,\phi_N(z))\,,\ee introduced in equation~\eqref{eq:phi-basis-def} consists of $M=N/2$ growing ($g$) and $M$ decaying ($d$) solutions at $0,\infty$ respectively. The structure of the connection problem can therefore be written as 
\be \begin{pmatrix}
d &
g
\end{pmatrix}^{(0)}
=\begin{pmatrix}
d &
g
\end{pmatrix}^{(\infty)}
\begin{pmatrix}
{E}_{dd} & {E}_{dg} \\
{E}_{gd} & {E}_{gg}
\end{pmatrix}.
\ee

We are looking for a solution $\phi(z) = \boldsymbol{\phi}^{(0)} \cdot C$ of (\ref{eq:operFGP}) where $C$ is a column vector of coefficients, such that $\phi(z)$ decays at both $0$ and $\infty$. If we choose $C$ to be such that $C_{i}=0$ for $i>M$, then clearly $\phi(z)$ decays at $0$. Moreover we can express $\phi(z)$ in terms of $\bm \phi^{(\infty)}$
\begin{equation}
    \phi(z) = \boldsymbol{\phi}^{(0)} C =\boldsymbol{\phi}^{(\infty)} E C\,,
\end{equation}
with $E= V^{-1}TV$. To ensure that $\phi(z)$ also decays at $0$, we must require that $(EC)_{i}=0$ for $i>M$ as well. Writing $C$ in the block form
\begin{equation}
    C= \begin{pmatrix}
        C_d \\0
    \end{pmatrix},
\end{equation}
where $C_d=(C_1,\dots, C_M)^T$, we understand that we need to impose $E_{gd}C_d=0$. Therefore, $E_{gd}$ has a kernel. Observing that $E_{gd}$ is the $M\times M$ bottom-left submatrix of $E$, with $M=N/2$, we conclude that
the quantization condition is 
\begin{equation}\label{eq:deteven}
   \det E[\swarrow]=0 \,.
\end{equation}
This determinant has been computed in equation~\eqref{det-formula-1}. We will now show that it is equivalent to 
\begin{equation} \label{eq:detE-as-FGP-even}
    \boxed{\det E[\swarrow] = \sum_{\bs{n}\in \mathcal{W}_{N}\cdot \bs{\lambda}_{N/2}} { 
   {\exp \left(\ri \pi\left(2 \bs{\eta}-{\bs \rho}\right)\cdot\bs{n}\right)} \over 
   {\prod_{\bs{\alpha}\in \Delta_{+}}\left(2\sin\left({\pi  \bs{\sigma}\cdot\bs{
   \alpha}}\right)\right)^{(\bs{n}\cdot{\bs \alpha})^2}} }}
\end{equation}

\begin{enumerate}
    \item First note that, since the Weyl group ${\cal W}_N$ for $\SU(N)$ is just the symmetric group $S_N$ it acts on the ${\bs e}_I$ by permuting the labels. Hence, recalling the expansion of the fundamental weight $\boldsymbol{\lambda}_{N/2}$ in terms of $\boldsymbol{e}_I$ (see \ref{eq:e-alpha-lambda}), we find that
   \be  \mathcal{W}_N\cdot {\bm \lambda}_{N/2}
=
\left\{
\sum_{i\in S} \bm e_i  \,\middle|\,  S\subset\{1,\dots,N\},\ \abs{S}=N/2
\right\}
.\ee This means that the vector $ \bm n=\sum_{i=1}^N n_i {\bm e}_i$  in the Weyl orbit of $\bm \lambda_{N/2}$ has
\be \label{eq:ntos} n_i=\begin{cases} 1\,, & i\in S \\ 0\,, & i\in S^c \end{cases}.\ee 
Any such permutation is clearly equivalent to a choice of subset $S\subset L=\{1,\dots,N\}$ with $\abs{S}=M$, since we only need to keep track of the positions of the $M=N/2$ pluses. 
Therefore the sum over the Weyl orbit in \eqref{eq:QCeven} is implemented automatically by the sum over $S$ in the Cauchy-Binet formula \eqref{det-formula-1}. 

\item Second, note that 
\begin{equation}\label{eq:etaprod}
    \bm\eta\cdot\bm n = \sum_{i=1}^N\eta_i\bm e_i\cdot\bm n  = \sum_{i\in S}\eta_i\, ,
\end{equation}
where we used equation~\eqref{eq:ntos}.
Therefore we have the relation
\begin{equation}
    \re^{2\pi\ii\bm\eta\cdot\bm n} = \prod_{i\in S}\zeta_i\,.
\end{equation}

\item  Next, let us look at the $\Sigma$-factors. We can rewrite
\begin{equation}\label{eq:sigmaterm}
    \frac{1}{\Sigma_i-\Sigma_j} = \frac{1}{e^{2\pi\ii\sigma_i} - e^{2\pi\ii\sigma_j}} = \frac{e^{-\pi\ii(\sigma_i+\sigma_j)}}{e^{\pi\ii(\sigma_i-\sigma_j)} - e^{\pi\ii(\sigma_j-\sigma_i)}} = \frac{e^{-\pi\ii(\sigma_i+\sigma_j)}}{2\ii\sin\pi(\sigma_i-\sigma_j)}\,.
\end{equation}
Consider now the numerator of the above. When performing the products in (\ref{det-formula-1}) each term $e^{-\pi \ii\sigma_i}$ will appear $M$ times, therefore the numerator reads
\begin{equation} \label{eq:sigmaterm2}
   \prod_{\substack{(i,j) \in \Sym(S\times S^c) \\ i>j}}e^{-\ri{\pi}(\sigma_i+\sigma_j)}=  \re^{-M\pi\ii\sum_{j=1}^N\sigma_j} = 1\,.
\end{equation}
Therefore we are left with 
\be   \label{eq:prod-Sigma}\prod_{\substack{(i,j) \in \Sym(S\times S^c) \\ i>j}} \frac{1}{\Sigma_i-\Sigma_j}=(-1)^{M}\prod_{\substack{(i,j) \in \Sym(S\times S^c) \\ i<j}}{1\over 2\ri \sin\pi\left(\sigma_i-\sigma_j\right)}\,.\ee

\item  
Coming back to \eqref{eq:QCeven}, we note that the exponent $(\bm n\cdot\bm\alpha)^2$ can only take the values $0$ or $1$. Indeed,
let us take $\bm\alpha\equiv\bm\alpha_{ij}=\bm e_i-\bm e_j$ with $i<j$ (see \ref{eq:positroot}).  Then
\begin{equation}
\bm n\cdot\bm\alpha
= \left(\sum_{k=1}^N n_k \bm e_k\right)\cdot(\bm e_i-\bm e_j)
= n_i-n_j = \begin{cases} \pm 1\,, & (i,j) \in \Sym(S\times S^c) \\
0\,, & \text{otherwise} 
\end{cases}.
\end{equation}
Moreover, for a given $\bm\alpha_{ij}$ the dot product appearing inside the sine in (\ref{eq:QCeven}) is simply
\begin{equation}
    \bm \sigma \cdot\bm\alpha_{ij} = \sum_{k=1}^N\sigma_k\bm e_k\cdot\bm\alpha_{ij} = \sum_{i=1}^N \sigma_k(\delta_{ki}-\delta_{kj}) = \sigma_i-\sigma_j\,,
\end{equation}

Therefore we find that\be\label{eq:sinsin}
\prod_{\substack{(i,j) \in \Sym(S\times S^c) \\ i<j}}{1\over 2\ri \sin\pi\left(\sigma_i-\sigma_j\right)} =\prod_{\bs{\alpha}\in \Delta_{+}}\left(2\ri\sin\left({\pi  \bs{\sigma}\cdot\bs{\alpha}}\right)\right)^{-(\bs{n}\cdot{\bs \alpha})^2}\,, \ee
\item 
Moreover we observe that
\begin{equation}
    \bm\rho = \sum_{i=1}^N\frac{N-2i+1}{2}\bm e_i \equiv \sum_{i=1}^N\rho_i\bm e_i\,,
\end{equation}
implies
\be \ba \bm\rho  \cdot \bm n= \sum_{i\in S}\rho_i= \sum_{i\in S} \frac{N-2i+1}{2}= M{2M+1\over 2}-\sum_{i\in S} i\,.\ea\ee
Hence
\be \label{eq:rho-match}
\re^{\ri \pi\,\bm{\rho}\cdot\bm n}
=
(-1)^{ \sum S}\,\re^{\ri \pi {M(2M+1)\over 2}}\,.\ee 

\item  Finally, we note that the signs $(-1)^{\sum M_2+ \sum S}$ from (\ref{det-formula-1}) cancel precisely with the products of $(-1)^{M^2}$ from (\ref{eq:prod-Sigma}), $(-1)^ {\sum S+M(2M+1)/2}$ from \eqref{eq:rho-match}, as well as $\re^{\ri  \pi/2  M^2}$ coming from the $\ri$ in front of the sine in \eqref{eq:sinsin}. 
\end{enumerate}
Hence we obtain \eqref{eq:detE-as-FGP-even} which concludes the proof of Part I of \autoref{th:thm1}. 

\subsection{Square-Integrable Solutions for \texorpdfstring{$N$}{N} Odd}\label{sec:proofodd}
When $N$ is odd, around each singularity, the bases $\bm \phi^{(0,\infty)}$ introduced in \autoref{sec:conn matrices} consist of  $(N-1)/2$ exponentially growing solutions ($g$),  $(N-1)/2$  exponentially decaying  solutions ($d$) and one oscillating solution with power law decay\footnote{In the $z$ variable.} ($o$) at $0,\infty$ respectively (see~\autoref{fig:roots-of-unity-ordered}). We will then schematically denote the two bases of solutions by $\boldsymbol{\phi}^{(0)}=(d\;o\;g)^{(0)}$ and $\boldsymbol{\phi}^{(\infty)}=(d\;o\;g)^{(\infty)}$. These are related by the connection matrix $E$, which takes the form
\be\label{eq:cceven} \begin{pmatrix}
d
&o
&g
\end{pmatrix}^{(0)}
= \begin{pmatrix}
d &
o &
g
\end{pmatrix}^{(\infty)}
\begin{pmatrix}
E_{dd} &E_{do} & E_{dg} \\
E_{od} &E_{oo}  & E_{og}\\
E_{gd} &E_{go}  & E_{gg}
\end{pmatrix}.\ee
In this case, the weakest conditions to obtain $L^2$ solutions can be organized as follows:
\begin{enumerate}
    \item We impose exponential decay at infinity, while allowing at 0 either exponential decay or power-law behavior.
\item We impose exponential decay at 0, while allowing at infinity either exponential decay or power-law behavior.
\item We allow both exponential or power-law decay at both 0 and infinity.
\end{enumerate}
As we will see below, the first two cases lead to resonance states and are directly related to \eqref{eq:QCodd} and \eqref{eq:QCodd2}. 
By contrast, the third case does not capture resonances. Indeed, it follows from equation \eqref{eq:cceven} that, under these boundary conditions, one can obtain an $L^2$ solution of equation~\eqref{eq:diffm} without imposing any quantization condition. The existence of such solutions is not surprising for operators which are not essentially self-adjoint \cite{Zettl2005, Caliceti1983,Caliceti,Maioli}. At the same time, it reflects the rich and subtle spectral structure exhibited by such operators, which certainly deserves further investigation.

\subsubsection{Case 1: Exponential Decay at \texorpdfstring{$+\infty$}{+∞}}
We look for a solution $\phi(z)$ of equation~\eqref{eq:operFGP} which, as a function of $z=\re^x$, decays exponentially at $z=\infty$ and either exponentially or with a power law at $z=0$. As in the $N$ even case, we enforce the latter requirement by letting $\phi(z) = \boldsymbol{\phi}^{(0)} \cdot C$, where 
$C$ is a column vector of coefficients such that $C_i = 0$ for $i >\tfrac{N+1}{2}$. Explicitly
\begin{equation}
    C=\begin{pmatrix}
        C_d \\
        C_o\\
        0
    \end{pmatrix}.
\end{equation}
Hence, upon writing $\phi(z) = \boldsymbol{\phi}^{(\infty)} \cdot EC$, it is manifest that requiring $\phi(z)$ to decay exponentially at $\infty$ amounts to imposing that $(EC)_i=0$ for $i\geq\frac{N+1}{2}$.
It follows that the submatrix
\be  E[\swarrow]=\begin{pmatrix}
E_{od} &E_{oo} \\
E_{gd} &E_{go}
\end{pmatrix},\ee 
namely the bottom-left ${N+1\over 2}\times {N+1\over 2}$  corner of $ E$, must have a kernel.
Hence, the quantization condition allowing for such a solution $\phi(z)$ is  \begin{equation}
   { \det E[\swarrow]=0}\,.
\end{equation}
By computing the determinant in \eqref{det-formula-1} for   $M={N+1\over 2}$ we now show that
\be\label{eq:detE-as-FGP-odd}
    \boxed{\det E[\swarrow] = \sum_{\bs{n}\in \mathcal{W}_{N}\cdot \bs{\lambda}_{N+1\over 2}} { 
   {\exp \left(\ri \pi\left(2 \bs{\eta}-{\bs \rho}+{\bs \sigma}\right)\cdot\bs{n}\right)} \over 
   {\prod_{\bs{\alpha}\in \Delta_{+}}\left(2\sin\left({\pi  \bs{\sigma}\cdot\bs{
   \alpha}}\right)\right)^{(\bs{n}\cdot{\bs \alpha})^2}} }}
\ee
To prove this, we first note that the sum over $S$ in \eqref{det-formula-1}  is now equivalent to a sum over the elements of the following Weyl orbit
  \be  \mathcal{W}_N\cdot {\bm \lambda}_{(N+1)/2}
=
\left\{
\sum_{i\in S} \bm e_i  \,\middle|\,  S\subset\{1,\dots,N\},\ \abs{S}=\frac{N+1}{2}
\right\}
.\ee   
 The difference with respect to the even \(N\) case is that \(S\) now contains \(M\) elements, whereas \(S^c\) contains \(M-1\) elements.
 This does not affect \eqref{eq:etaprod} and we still have
\be e^{2\pi\ii\bm\eta\cdot\bm n} = \prod_{s\in S}\zeta_s\,.\ee
However, a crucial difference now arises when evaluating the $\Sigma$-dependent terms in \eqref{eq:sigmaterm}.
As before, we have
\begin{equation}
    \frac{1}{\Sigma_i-\Sigma_j} = \frac{e^{-\ri{\pi}(\sigma_i+\sigma_j)}}{2\ii\sin{\pi}(\sigma_i-\sigma_j)}\,.
\end{equation}
However, in the $N$ even case the exponentials canceled, see equation~\eqref{eq:sigmaterm2}. By contrast now, considering that in the product over $(i,j) \in \Sym(S\times S^c)$, every factor $e^{-i\pi\sigma_i}$ with $i \in S$ appears as many times as there are elements in $S^c$ (and vice versa), we have 
\be\label{eq:change}
  \prod_{\substack{(i,j) \in \Sym(S\times S^c) \\ i>j}}e^{-\ri{\pi}(\sigma_i+\sigma_j)}= e^{-\ri{\pi}\left( M\sum_{j\in S^c}\sigma_j + (M-1)\sum_{i\in S}\sigma_i \right)} = \prod_{i\in S} \re^{\ri{\pi}\sigma_i}\,.
\ee
having used as before the fact that $\sum_{i\in S}\sigma_i +\sum_{j\in S^c}\sigma_j =\sum_{i=1}^N\sigma_i=0$.

Hence, since $\bm \sigma\cdot\bm n = \sum_{s\in S}\sigma_s$, we now find that
\be \label{eq:sigma-to-sin-odd}
\prod_{\substack{(i,j) \in \Sym(S\times S^c) \\ i>j}} \frac{1}{\Sigma_i-\Sigma_j} =\re^{{\pi \ri \boldsymbol{\sigma}\cdot {\bs n}}} \prod_{\bs{\alpha}\in \Delta_{+}}{1\over \left(2\ri\sin\left({\pi  \bs{\sigma}\cdot\bs{\alpha}}\right)\right)^{(\bs{n}\cdot{\bs \alpha})^2}}\,, \ee
Finally, we find that \eqref{eq:rho-match} becomes
\be \label{eq:rho-match2}
\re^{\ri \pi\,\bm{\rho}\cdot\bm n}
=
(-1)^{\sum S}\,\re^{\ri \pi {M ^2}}.\ee 
Also in this case, the signs on both sides cancel and we get \eqref{eq:detE-as-FGP-odd}, which concludes the proof of Part II, case 1 of \autoref{th:thm1}.

\subsubsection{Case 2: Exponential Decay at \texorpdfstring{$-\infty$}{-∞}}
We look for a solution to (\ref{eq:operFGP}) that, as a function of $z$, decays exponentially at $z=0$, allowing either exponential or power-law decay as $z \to \infty$.  One may proceed as in the previous sections; alternatively, one can observe that exchanging \( 0 \) and \( \infty \) corresponds to the transformation \( E \to E^{-1} \). This, in turn, is equivalent to \( \bm\eta \to -\bm\eta \), since \( E^{-1} = V^{-1} T^{-1} V \) and \( T^{-1} = \operatorname{diag}(\zeta_1^{-1}, \dots, \zeta_N^{-1}) \). Hence, we obtain \eqref{eq:QCodd2}, which concludes the proof of Part II, case 2 of \autoref{th:thm1}.

\section{Proof of Corollary}\label{sec:corollproof}

We now want to prove \autoref{coroll}. For that we first note that
the relation between \eqref{eq:diffm} and \eqref{eq:operFGP} is encoded by the Fourier transform \eqref{eq:Fourier-conventions}.
This means that, if $\phi(x)$ is a solution to \eqref{eq:operFGP} satisfying the assumptions of \autoref{th:thm1}, the Fourier integral
\be\label{eq:ft}
\psi(y)=\frac{1}{\sqrt{2\pi\hbar}}\int_{\BR}\dd x\, \re^{-{\ri\over \hbar}xy}\phi(x)\,,
\ee
exists and $\psi(y)$ solves equation \eqref{eq:diffm}.\footnote{The case $N=3$ is slightly exceptional, and will be dealt with at the end of this section.}
The latter point is slightly more subtle than it appears.
In order to prove that the Fourier transform indeed solves~\eqref{eq:diffm}, one needs to show that all of the required manipulations can be performed inside the integral.
For $N$ even, this is automatic, however for $N$ odd, neither $e^{\pm x}\phi(x)$ or $\partial_x^N\phi(x)$ are bounded at $\pm\infty$ if we assume the oscillatory asymptotics~\eqref{eq:oscillating-sol}.
To remedy this, we can first shift the integration contour slightly down by Cauchy's theorem,\footnote{To be precise, the vanishing of the vertical contribution at infinity follows from uniform decay on closed strips, which in turn follows from Borel summability~\cite[Theorem 7.2]{Sauzin:2014qzt}.} such that the asymptotics of the integrand become (w.l.o.g. we consider oscillatory behavior at $+\infty$)
\begin{equation} \label{eq:osc-to-decay}
    \phi(x-\ii\delta) \sim \re^{-\ii N{\Lambda\over \hbar}\re^\frac{x-\ii\delta}{N}} \implies \abs{\phi(x-\ii\delta)} \sim \re^{-N {\Lambda\over \hbar}\sin\frac{\delta}{N}\re^\frac{x}{N}}\,,\qquad x\to+\infty\,,
\end{equation}
which decays faster than any exponential for $0<\delta<N\pi$.\footnote{More precisely we need $0<\delta<\pi$ so that the contribution from the next-slowest decaying solution does not become growing, cf.~\autoref{fig:roots-of-unity-ordered}.} 
We can then compute 
\begin{align} \label{eq:oper-implies-Baxter}
    \left( (-1)^{N+1}u_N - V_N(y) \right)\psi(y) &= \frac{1}{\sqrt{2\pi\hbar}}\int_\BR\dd x\,\re^{-\frac{\ii}{\hbar}(x-\ii\delta)y}\left( (-1)^{N+1}u_N - V_N(-\ii\hbar\partial_x) \right)\phi(x-\ii\delta) \nonumber\\
    &= \frac{1}{\sqrt{2\pi\hbar}}\int_\BR\dd x\, \re^{-\frac{\ii}{\hbar}(x-\ii\delta)y}\Lambda^N\left( \re^{x-\ii\delta}+\re^{-(x-\ii\delta)} \right)\phi(x-\ii\delta) \nonumber\\
    &=\frac{\Lambda^N}{\sqrt{2\pi\hbar}}\int_\BR\dd x \left( \re^{-\frac{\ii}{\hbar}(x-\ii\delta)(y+\ii\hbar)} + \re^{-\frac{\ii}{\hbar}(x-\ii\delta)(y-\ii\hbar)} \right)\phi(x-\ii\delta) \nonumber\\
    &= \Lambda^N(\psi(y+\ii\hbar) + \psi(y-\ii\hbar))\,,
\end{align}
where we used equation~\eqref{eq:operFGP} and the strong decay of the integrand.

For the converse statement we have
\be\label{eq:ftinv}
\phi(x)=\frac{1}{\sqrt{2\pi\hbar}}\int_{\BR}\dd y\, \re^{{\ri\over \hbar}xy}\psi(y)\,,
\ee
and we must make use of a similar trick, as we do not expect either of the combinations $\psi(y+\ii\hbar)+\psi(y-\ii\hbar)$ or $V_N(y)\psi(y)$ to be integrable on the real line (for $N$ odd).
However since (in both the even and odd cases) we require the integrand to be in $L^1\cap L^2$ on a (semi-)infinite strip, we may simply shift the integration contour such that it is centered on a closed strip of width $2\hbar$ contained in the larger strip.\footnote{Here the vanishing of the vertical contribution at infinity follows from the assumption of uniform boundedness and decay on closed strips.}
This way $\psi(y+\ii\hbar)+\psi(y-\ii\hbar)$ is clearly in $L^1\cap L^2$, as is $V_N(y)\psi(y)$ thanks to equation~\eqref{eq:diffm}.
From there it follows that the Fourier transform of $\psi(y)$ solves equation~\eqref{eq:operFGP}.

In the following we make extensive use of the Paley-Wiener theorem relating analytic properties of a function to the decay properties of its Fourier transform, and vice versa.
For convenience, we recall the relevant statements here, which are adapted from~\cite[Theorem IX.13-14]{reed1975ii}
\begin{theorem} \label{th:PW} (Paley-Wiener)
    \paragraph{Part I:}
    Let $-a_-,a_+\in\BR_{>0}$ (possibly infinite), and $\phi$ be a function for which $\re^\frac{xb}{\hbar}\phi\in L^1(\BR)\cap L^2(\BR)$ for all $a_-<b<a_+$.
    Then the Fourier transform
    \begin{equation} \label{eq:psi-from-phi}
        \psi(y) = \frac{1}{\sqrt{2\pi\hbar}}\int_\BR\dd x\, \re^{-\frac{\ii}{\hbar}xy}\phi(x)\,,
    \end{equation}
    exists and admits an analytic continuation to the strip $\{y\in\BC\,|\, a_-<\Im y<a_+\}$ such that $\psi(\bullet+\ii b)\in L^2(\BR)$ for all $a_-<b<a_+$.
    \paragraph{Part II:} 
    Let $-a_-,a_+\in\BR_{>0}$ (possibly infinite), and $\psi$ be a holomorphic function such that $\psi(\bullet+\ii b)\in L^1(\BR)\cap L^2(\BR)$ for all $a_-<b<a_+$.
    Then the inverse Fourier transform 
    \begin{equation} \label{eq:phi-from-psi}
        \phi(x) = \frac{1}{\sqrt{2\pi\hbar}}\int_\BR\dd y\,\re^{\frac{\ii}{\hbar}xy}\psi(y)\,,
    \end{equation}
    exists and, if it is independent of the choice of integration contour in the strip,\footnote{E.g. by uniform boundedness and decay on closed strips.} then it is bounded by $C_b\re^{-\frac{bx}{\hbar}}$ for some constant $C_b$ for all $a_-<b<a_+$.
    In particular this implies that $\phi$ decays faster than $\re^{-\frac{b_+}{\hbar}x}$ as $x\to+\infty$ for all $b_+<a_+$, and faster than $\re^{-\frac{\abs{b_-}}{\hbar}\abs{x}}$ as $x\to-\infty$ for all $0>b_->a_-$.
\end{theorem}
The proof of Part I follows by evaluating the integral at $y+\ii b$, while for Part II it follows from shifting the integration contour to $\Im y=b$.
Similar statements hold when replacing $\re^{\pm\frac{\ii}{\hbar}xy}$ with $\re^{\mp\frac{\ii}{\hbar}xy}$ in the integral by replacing the strip with its complex conjugate, that is $a_\pm\to-a_\mp$.

\subsection{Square-Integrable Solutions for \texorpdfstring{$N$}{N} Even}
Consider first the case where $N$ is even.

Let us assume \eqref{eq:QCeven} holds.
Part I of \autoref{th:thm1} implies that, if the condition \eqref{eq:QCeven} is satisfied, then the corresponding eigenfunctions of \eqref{eq:operFGP} belong to \(L^2(\BR)\). From \cite{Baerman:2025uzv} we know that these functions decay super-exponentially as $x \to \pm \infty$ on a strip around the real axis as long as $ \abs{\Im(x)}<\pi$.

This can be understood pictorially by observing~\autoref{fig:roots-of-unity-ordered}. Shifting the imaginary part of $x$ by $x\rightarrow x+\ii\theta$ is equivalent to rotating the roots of unity by $\re^{\ii \theta/N}$. 
It follows that the set of $N/2$ decaying solutions of (\ref{eq:operFGP}) remains decaying under the shift $x\rightarrow x+\ii\theta$ if the roots of unity lying in the green-shaded region remain inside the same region after the rotation by $e^{\ii \theta/N}$. It suffices then to realize that the minimum angle between a root of unity $\alpha_k$ and the imaginary axis in \autoref{fig:roots-of-unity-ordered} is $\pi/N$, hence $\abs{\theta}<\pi$.

Thus, there exist $a,b>0$ such that
\be\label{eq:asymphiin}
|\phi(x)|\lesssim \re^{-a \re^{b|{\rm Re}(x)|}}, \qquad \abs{x}\to\infty\,,\quad \abs{{\rm{Im}} (x)}={\rm{const.}} < \pi\,.
\ee
This decay is in particular stronger than any exponential, so that \(\re^{c|x|}\phi(x)\in L^1\cap L^2\) for all $c>0$ and $\abs{\Im(x)}<\pi$.
By Part I of \autoref{th:PW}, $\psi(y)$ is therefore entire and in $L^2$ along any line parallel to the real axis.
Furthermore, by Part II, it decays faster  then  $\re^{-\frac{a}{\hbar}\abs{y}}$ for any $0<a<\pi$, which is in agreement with the results of~\cite{Francois:2025nmq}.
In particular, we obtain the bound 
\begin{equation}   
    \abs{\psi(u+\ii v)}\leq\frac{1}{\sqrt{2\pi\hbar}}\re^{\frac{b}{\hbar}u}\left\| \phi(x+\ii b)\max\!\left( \re^{\alpha_-x/\hbar},\re^{\alpha_+x/\hbar} \right)\! \right\|_{L_1},
\end{equation}
for every $-\pi<b<\pi$, which is uniform on any closed strip $\alpha_-\leq v\leq\alpha_+$.

Let us now prove the inverse. Assume $\psi(y)$ is such that for any fixed $\mathrm{Im}(y)$, its restriction to the line $\BR + \ri\,\mathrm{Im}(y)$  is in $L^1\cap L^2$ and decays exponentially along any line parallel to the real axis, and the contour shift is allowed. 
Then by the Plancherel Theorem and the argument presented at the beginning of this section, its Fourier transform exists, is in $L^2$, and solves equation~\eqref{eq:operFGP}. Hence, by Part I of \autoref{th:thm1}, we get that \eqref{eq:QCeven} is satisfied. 

\subsection{Square-Integrable Solutions for \texorpdfstring{$N$}{N} Odd}

Let us now consider the case where $N$ is odd.
\subsubsection{Case 1: Square Integrability in the Upper Half-Plane}\label{sec:corollodd1}

Let us assume \eqref{eq:QCodd} holds. 
Part II, case 1 of  \autoref{th:thm1} implies that  the corresponding solutions of \eqref{eq:operFGP} belong to $L^1(\mathbb{R})\cap L^2(\BR)$. By assumption they exhibit super-exponential decay as $x \to +\infty$, while they may decay either exponentially 
or super-exponentially as $x \to -\infty$. Moreover, by \cite{Baerman:2025uzv} -- see equations \eqref{eq:phi-basis-def} and \eqref{eq:chi-asymptotics} -- when analytically continued along lines parallel to the real axis, they remain in $L^2(\mathbb{R}+\ii\Im x)\cap L^1(\BR +\ii\Im x)$ provided that $-\pi < {\rm Im}\,x < 0$. Specifically, if a solution decays super-exponentially on the real line, then its analytic continuation to the strip \( -\pi < \operatorname{Im} x < 0 \) preserves this rate of decay; in contrast, if the solution is only exponentially decaying, its analytic continuation enhances the decay to a super-exponential rate.

As in the even case, this can be understood pictorially from \autoref{fig:roots-of-unity-ordered} (see also equation~\eqref{eq:oscillating-sol}) by observing that rotations by $\re^{\ii\theta/N}$ with $-\pi<\theta<0$ maintain the $N$-th roots of unity in the green-shaded region: note how the root of unity sitting on the imaginary axis will be rotated inside the green region, thus enhancing the power-law decay to a super-exponential one. 

It then follows by Part II of \autoref{th:PW} that 
\begin{equation}
|\psi(u + \ii v)| \le C(v,b)\, e^{-\frac{b}{\hbar} u}\,, \qquad u \to +\infty\,,\quad \forall\, 0<b< \pi\,.
\end{equation}
However, as \( u \to -\infty \), the same argument no longer applies. Indeed, since we must have $-\pi < {\rm Im}\,x < 0$, shifting the contour produces an exponentially growing factor rather than a decaying one. Nevertheless, we can argue for a weaker decay by considering the subleading behavior of $\phi(x)$. 
From equation~\eqref{eq:oscillating-sol} we see that for $x\to-\infty$, we still have exponential decay of the form
\begin{equation}
    \abs{\phi(x)} \sim e^{-\frac{N-1}{2N}\abs{x}}\,,\qquad x\to-\infty\,,
\end{equation}
which ensures that $ e^{\frac{v}{\hbar} x }\phi(x)$ lies in $L^1(\BR)\cap L^2(\BR)$.
Thus, by Part I of \autoref{th:PW}, the Fourier transform $\psi(y)$ given by~\eqref{eq:psi-from-phi} is analytic and in $L^2(\BR+\ii v)$ on the strip 
\begin{equation}
\label{eq:decaycoroll}
\operatorname{Im} y = v > -\hbar\left(\frac{1}{2} - \frac{1}{2N}\right).
\end{equation}
To see that it is also analytic on the rest of the complex plane, we employ the same contour shift as in equation~\eqref{eq:oper-implies-Baxter} to obtain
\begin{equation}
    \psi(y) = \frac{1}{\sqrt{2\pi\hbar}}\int_\BR\dd x\, \re^{-\frac{\ii}{\hbar}xy}\phi(x) = \frac{\re^{-\frac{\delta y}{\hbar}}}{\sqrt{2\pi\hbar}}\int_\BR\dd x\,\re^{-\frac{\ii}{\hbar}xy}\phi(x-\ii\delta)\,.
\end{equation}
Since the integrand now decays super-exponentially as~\eqref{eq:osc-to-decay}, $\psi(y)$ is entire.
Note however that it is not necessarily in $L^2(\BR+\ii v)$ for $v\in\BR\setminus\eqref{eq:decaycoroll}$ due to the exponential prefactor.
Instead it is allowed to have sub-exponential growth as $y\to-\infty$ (since $\delta>0$ may be taken arbitrarily small), in agreement with the polynomial growth found in~\cite{Francois:2025nmq}.

It remains to argue that \( \psi(u + \ii v) \) is in $L^1 (\BR +\ii v)$ in some semi-infinite strip.  For this we note that
\be y \psi(y)\propto 
\int_{\mathbb{R}} \mathrm{d}x\, 
\re^{-\frac{\ii}{\hbar}x y}\,\partial_x\phi(x) \,,
\ee
where the integrand on the right hand side is again in  $L^1\cap L^2$ provided that 
\be\label{eq:decaybad} \Im y> -{\hbar\over 2}\left(1-{3\over N}\right).\ee
Hence by the Cauchy-Schwarz-Bunyakovsky  inequality we get
\be \|{\psi(u + \ri v)}\|_{L^1}\leq  \|(1+|u|){\psi(u + \ri v)}\|_{L^2} \left\|{1\over |u|+1} \right\|_{L^2} <\infty \,.\ee
Lastly, uniform boundedness on closed strips follows from
\begin{equation}
    \abs{\psi(u+\ii v)} \leq \frac{1}{\sqrt{2\pi\hbar}} \left\| \phi(x)\max\!\left( \re^{\alpha_-x/\hbar}, \re^{\alpha_+x/\hbar} \right)\!\right\|_{L_1},
\end{equation}
which holds for $\alpha_-\leq v\leq\alpha_+$ and $\alpha_\pm>-\frac{\hbar}{2}\left(1-\frac{3}{N}\right)$.

Let us now prove the converse. Suppose there exists a non-trivial entire solution to \eqref{eq:diffm} which lies in $L^2$ in the semi-infinite strip \eqref{eq:decaycoroll} and in $L^1$ for
\be \label{eq:L1-strip} \Im y > -{\hbar\over 2}\left(1-{3\over N}\right).\ee
When $N>3$, its Fourier transform belongs to $L^2(\BR)$ and decays faster than any exponential at $+\infty$ by Part II of \autoref{th:PW}. Following the argument presented at the beginning of the section its Fourier transform solves equation \eqref{eq:operFGP}.
When $N=3$, the real line is not included in the half-plane~\eqref{eq:L1-strip}, so the ordinary Fourier transform may not be well-defined.
Instead we may consider the same integral along a shifted contour $\Im y=v>0$, such that the integral is convergent.
Then, analogously to the case $N>3$, the integral solves equation~\eqref{eq:operFGP} and decays faster than any exponential at $+\infty$, at the cost of only being bounded by $\abs{\phi(x)}\leq (2\pi\hbar)^{-1/2}\|\psi(u+\ii v)\|_{L_1}\re^{-\frac{v}{\hbar}x}$ instead of just a constant.
However, as can be seen from the asymptotics~\eqref{eq:chi-asymptotics}, equation~\eqref{eq:operFGP} does not possess any solutions which grow only exponentially, therefore $\phi(x)$ must be bounded.
Hence Part II, case 1 of \autoref{th:thm1} implies that \eqref{eq:QCodd} is satisfied for any odd $N$.

\subsubsection{Case 2: Square Integrability in the Lower Half-Plane}
This is analogous to \autoref{sec:corollodd1}. However, since now the decay is super-exponential as $x \to -\infty$, while as $x \to +\infty$ it can be exponential, the condition \eqref{eq:decaycoroll} is replaced by
\be\label{eq:decaycoroll2}
 \Im y < \hbar\left(\frac{1}{2} - \frac{1}{2N}\right),
\ee
while \eqref{eq:decaybad} becomes
\be  \Im y< {\hbar\over 2}\left(1-{3\over N}\right).\ee

\subsubsection{Comment on Continuous Spectrum}\label{sec:cont}
Let us consider the third possibility discussed in \autoref{sec:proofodd}, where only exponential decay as \( x \to \pm\infty \) is assumed and we obtain an \( L^2(\mathbb{R}) \) solution without the need to impose any quantization condition. On the Fourier transform side, this decay implies that, as \( \operatorname{Re} y \to -\infty \), the analytic continuation of the Fourier transform preserves square integrability only within a restricted strip:
\be  \label{eq:immcont}-\hbar\left(\frac{1}{2} - \frac{1}{2N}\right) < \Im y <  \hbar\left(\frac{1}{2} - \frac{1}{2N}\right). \ee

\section{Outlook}\label{sec:outlook}

In this work, based on  \cite{Baerman:2025uzv, Francois:2025nmq}, we have demonstrated the quantization conditions proposed in \cite{Grassi:2018bci} and summarized in \autoref{th:thm1}. Such conditions can be interpreted as a minimal requirement ensuring the existence of $L^2$ solutions of the $N$-th order differential equation \eqref{eq:operFGP}.\footnote{For $N$ odd there is one even weaker condition leading to a continuous spectrum as discussed above.} Equivalently, as follows from~\autoref{coroll}, such conditions give normalizable eigenfunctions of the finite-difference operator~\eqref{eq:diffm} with respect to the Lebesgue measure $\dd y$. However, the richness of higher-order differential equations allows one to impose stronger constraints. For instance, in \cite{Baerman:2025uzv} a spectral problem was considered in which the solutions are required to be square-integrable with exponential decay of the form
\begin{equation}
    \chi(x) \sim \re^{-N{\Lambda\over \hbar}\re^\frac{\abs{x}}{N}}\,,\qquad x\to\pm\infty\,.
\end{equation}
This corresponds to selecting solutions that are maximally decaying both at $-\infty$ and at $+\infty$. In this setting, a solution to \eqref{eq:operFGP} exists only if $N-1$ independent quantization conditions are imposed.
From a quantum mechanical perspective, these again correspond to $L^2$ normalizable solutions of the Baxter equation~\eqref{eq:baxter}, but with respect to the Sklyanin measure $\dd\mu_\mathrm{Skl}(y)$, which is the natural choice for the Toda chain~\cite{Sklyanin1985}.

Our results suggest the existence of a broader family of quantization conditions, labeled by an integer $K$, which interpolate between the case of \cite{Grassi:2018bci} (corresponding to $K=1$) and that of \cite{Baerman:2025uzv} (corresponding to $K=N-1$). From the perspective of the Fourier-transformed  difference operator \eqref{eq:diffm}, we expect that the  eigenfunctions for all these spectral problems will arise from the “parent” eigenfunction constructed in \cite{Francois:2025nmq}, upon imposing progressively stronger decay conditions.

A related and intriguing feature of equation 
\eqref{eq:diffm} for $N>2$ is the existence of special points where tunneling is suppressed \cite{Grassi:2018bci}. This phenomenon has recently been revisited from the perspective of WKB analysis and resurgence in \cite{Gu:2026xgp}. It would be interesting to understand how these special points are encoded in the connection matrix introduced in \cite{Baerman:2025uzv}, and whether this framework could lead to a complete classification of them.

More generally, the spectral theory of the operators with $N$ odd appears to be substantially richer than in the even case and deserves further investigation. In particular, it would be important to achieve a sharper understanding of the full spectral decomposition, including the continuous spectrum discussed above.

We hope to return to these questions in future work.

\appendix

\section{Conventions for the \texorpdfstring{$\mathfrak{su}(N)$}{su(N)} Root System}
\label{app:GT}
Our conventions follow those of \cite{Francois:2025nmq}, which we briefly summarize here for completeness and to ensure a self-consistent presentation. Our conventions for the root system of $\mathfrak{su}(N)$ are as follows:  
\begin{equation}\label{eq:latticedef}
    \begin{cases}
        \cbr{\bs{b}_i}_{i \in \cbr{1, \cdots, N}}  & \text{standard Euclidean orthonormal basis;}
        \\
        \cbr{\bs{e}_i}_{i \in \cbr{1, \cdots, N}}  & \text{weights of the fundamental representation;}
        \\
        \cbr{\bs{\lambda}_k}_{k \in \cbr{1, \cdots, N-1}} & \text{fundamental weights, basis of the weight lattice;}
        \\
        \cbr{\bs{\alpha}_\ell}_{\ell \in \cbr{1, \cdots, N-1}}  & \text{simple roots, dual to the fundamental weights;}
    \end{cases}
\end{equation}
where 
\be  \bs{e}_i \cdot \bs{e}_j = \delta_{i, j} - \frac{1}{N} \, ,\qquad \sum_{i=1}^N \bs{e}_i=0     \, . \ee
We also recall that
\begin{equation}\label{eq:e-alpha-lambda}
\bs{e}_i = \bs{b}_i - \frac{1}{N} \sum_{j = 1}^N \bs{b}_j
    \, ,
    \qquad
    \bs{\alpha}_k = \bs{e}_k - \bs{e}_{k+1} = \bs{b}_k - \bs{b}_{k+1}
    \, ,\qquad \bs{\lambda}_k = \sum_{i = 1}^k \bs{e}_i    \, .
\end{equation}
We define the set of positive roots as:
\begin{equation}
    \label{eq:positroot}\Delta^+=\left\{\bs{\alpha}_{k, \ell} = \bs{e}_k - \bs{e}_\ell\, \Big| \,  1 \leqslant k< \ell \leqslant N \right\}.\end{equation}
The  Weyl group  $\mathcal{W}_N$ is isomorphic to the symmetric group $S_N$ and acts on the weight lattice by permuting the indices of $e_i$. The Weyl orbit of an element 
\be {\bs v}=\sum_{i=1}^N v_i e_i\,, \ee is denoted as
\begin{equation}
    \mathcal{W}_N\cdot v=\left\{ w({\bs v})\,|\, w\in \mathcal{W}_N\right\}.
\end{equation}

\section{Floquet Solutions and Associated Monodromy}\label{app:floquet}
For completeness, let us briefly review the construction of the Floquet solutions of equation~\eqref{eq:operBRT} constructed in \cite{Baerman:2025uzv}.
One first introduces the infinite determinants $K_\pm$ as
\begin{equation}\label{eq:Kpm}
K_+(\lambda,\Lambda, \bm{u}) =
\det
\begin{pmatrix}
1 & \dfrac{1}{t(\lambda + \ii\hbar)} & 0 & 0 & \cdots \\
\dfrac{\Lambda^{2N}}{t(\lambda + 2\ii\hbar)} & 1 & \dfrac{1}{t(\lambda + 2\ii\hbar)} & 0 & \cdots \\
0 & \dfrac{\Lambda^{2N}}{t(\lambda + 3\ii\hbar)} & 1 & \dfrac{1}{t(\lambda + 3\ii\hbar)} & \cdots \\
0 & 0 & \dfrac{\Lambda^{2N}}{t(\lambda + 4\ii\hbar)} & 1 & \cdots \\
\vdots & \vdots & \vdots & \vdots & \ddots
\end{pmatrix}.
\end{equation}
   where \begin{equation}
t(\lambda)
= \lambda^N + \sum_{k=0}^{N-2} \lambda^k\,(-1)^{N-k} u_{N-k}
= \prod_{k=1}^N (\lambda - \tau_k)\, .
\end{equation} and $K_-(\lambda,\Lambda,\bm u)=\overline{K_+(\overline{\lambda},\overline{\Lambda},\overline{\bm u}) }$. Then the functions
\begin{subequations} \label{eq:qdef}
\begin{align}
Q^+_{\bm\tau}(\lambda)
&=
\left(\frac{\hbar}{\Lambda}\right)^{\frac{\ii N\lambda}{\hbar}}
\frac{e^{-\frac{N\pi\lambda}{\hbar}}
K_+(\lambda, \Lambda,\bm u)}{\prod_{k=1}^N\Gamma \!\left(1 - \frac{\ii}{\hbar}(\lambda - \tau_k)\right)}\,,\\
Q^-_{\bm\tau}(\lambda)
&=
\left(\frac{\Lambda}{\hbar}\right)^{\frac{\ii N\lambda}{\hbar}}
\frac{e^{-\frac{N\pi\lambda}{\hbar}}
K_-(\lambda, \Lambda,\bm u)}{\prod_{k=1}^N\Gamma \!\left(1 + \frac{\ii}{\hbar}(\lambda - \tau_k)\right)}\,,
\end{align}
\end{subequations}
are solutions to the Baxter equation (\ref{eq:baxter}), as first shown in \cite{Gutzwiller1980,Gutzwiller1981}.
Note that the poles of $K_\pm$ in $\lambda$ are precisely canceled by those of the $\Gamma$ functions in \eqref{eq:qdef}, making $Q^\pm_{\tau}(\lambda)$ entire \cite{Gutzwiller1980,Gutzwiller1981,PasquierGaudin1992}.

The determinant \eqref{eq:Kpm} exists and is absolutely convergent away from its poles for any value of $\Lambda$. This follows from the general criterion for infinite determinants, which states that convergence is ensured provided that the product of the diagonal elements and the sum of the off-diagonal elements are absolutely convergent (see \cite[paragraph 2.82 and Example 2.8.1]{Whittaker_Watson_1996}).

The Floquet solutions to the equation (\ref{eq:operBRT}) are then defined as \cite[Prop.~4]{Baerman:2025uzv}
\begin{align}\label{eq:FtoQ}
F^{(0)}_j(z)
&=
\sum_{n \in \mathbb{Z}}
Q^+_{\bm\tau}\!\left(-\ri\hbar \sigma_j  - \ri\hbar n\right)\,
z^{\sigma_j + n},
\qquad j = 1,\dots,N,
\\[6pt]
\label{eq:FtoQ1}F^{(\infty)}_j(z)
&=
\sum_{n \in \mathbb{Z}}
Q^-_{\bm\tau}\!\left(-\ri\hbar \sigma_j  - i\hbar n\right)\,
z^{\sigma_j + n},
\qquad j = 1,\dots,N,
\end{align}
where the Floquet exponents $\sigma_j$  are the $N$ solutions (up to integer shifts) of the quantum Wronskian equation
\be \label{eq:Qwronsk} W(-\ri\hbar \sigma_j)=0\,,\ee
where
\be W(\lambda)=Q^+_{\bm\tau}(\lambda)\,Q^-_{\bm\tau}(\lambda+\ii\hbar)
-
Q^-_{\bm\tau}(\lambda)\,Q^+_{\bm\tau}(\lambda+\ii\hbar)\,.\ee
One can show that $W(\lambda)=(-1)^N W(\lambda+\ri\hbar)$ and that \eqref{eq:Qwronsk} generically admits $N$ solutions in the strip $\left\{\lambda\in\BC\,\middle|\,\abs{\Im \lambda}<\frac{\hbar}{2}\right\}$~\cite{Gutzwiller1980,Gutzwiller1981,PasquierGaudin1992} such that
\be \sum_{j=1}^N \sigma_j=0~.\ee
It is easy to show (see \cite{Baerman:2025uzv}) that the fact that $Q^{\pm}_{\bm\tau}$ are solutions to the Baxter equation (\ref{eq:baxter}) ensures that the Floquet series (\ref{eq:FtoQ}, \ref{eq:FtoQ1}) are, at least formally, solutions to (\ref{eq:operBRT}). Furthermore, the fact that $\sigma_j$ satisfies \eqref{eq:Qwronsk} ensures that the series \eqref{eq:FtoQ} is absolutely convergent for $0<|z|<\infty$ \cite[Prop.~4]{Baerman:2025uzv}. It then follows that the sets of functions $\{F^{(0,\infty)}_j(z)\}_{j=1}^N$ form two bases of solutions to the differential equation \eqref{eq:operBRT}, whose elements are eigenvectors of the monodromy with identical eigenvalues, namely
\be F^{(0, \infty)}_j(\re^{2\pi \ri }z) = F^{(0, \infty)}_j(z)\re^{2\pi \ri \sigma_j}\,.\ee
The two Floquet bases must then be related by a diagonal change of basis, i.e.
\be  {F^{(0)}_j(z)}=F^{(\infty)}_j(z)\re^{2\pi \ri\eta_j}. \ee
Having argued that the Laurent series \eqref{eq:FtoQ} are convergent, we obtain the following explicit relation between $\eta_i$ and $\sigma_j$ \cite{Baerman:2025uzv}
\begin{equation}\label{eq:eta-sigma}
\re^{2\pi\ii \eta_j}
=
\frac{Q^+_{\bm\tau}(-\ri\hbar \sigma_j)}{Q^-_{\bm\tau}(-\ri\hbar \sigma_j)}\eqqcolon\zeta_j\, .
\end{equation}
As a final remark, note that solving \eqref{eq:Qwronsk} gives an explicit relation between the monodromy parameters $\sigma_j$  and the parameters of the differential equation, $u_k$ (or equivalently $\tau_k$). Locally, this relation can be inverted  explicitly to obtain $u_k$ as a function of $\sigma_i$ \cite{Kozlowski:2010tv, Meneghelli:2013tia,Flume:2004rp, Bullimore:2014awa, Gaiotto:2014ina, Grassi:2018bci}. 
For example in the small $\Lambda$ regime, one finds 
\begin{equation}
\label{eq:utoam}
u_k = s_{(1^k)}(-\ri \hbar \sigma_j) + \mathcal{O}(\Lambda^N)\,,
\end{equation}
where $s_R$ denotes the Schur polynomial associated with the Young tableau $R$, and $(1^k)$ corresponds to the tableau with $k$ rows and one box per row. Equation \eqref{eq:utoam} represents the quantum version of the Matone relations, originally discovered in \cite{Matone:1995rx}.
However, explicit expressions of such relations are not required for our purposes.

\bibliographystyle{ytphys}
\bibliography{biblio}

\end{document}